\documentclass[sort,AMA,LATO2COL]{WileyNJD-v2}
\usepackage{moreverb}
\usepackage{amssymb}
\usepackage{amsmath}
\usepackage{array}
\newcolumntype{C}[1]{>{\centering\arraybackslash}p{#1}}

\usepackage{bibspacing} 		
\usepackage{boldline}			
\usepackage{color}
\usepackage[inline]{enumitem}
\setlist{nolistsep}
\usepackage[keeplastbox]{flushend}
\usepackage{graphicx}
\usepackage{multibib}
\usepackage{multicol}
\usepackage{multirow}
\usepackage{rotating}
\usepackage[caption=false]{subfig}

\usepackage{pifont}
\usepackage{setspace}

\newcolumntype{P}[1]{>{\centering\arraybackslash}p{#1}}

\graphicspath{{./images/}{./images/Performance/}}

\newcommand\BibTeX{{\rmfamily B\kern-.05em \textsc{i\kern-.025em b}\kern-.08em
T\kern-.1667em\lower.7ex\hbox{E}\kern-.125emX}}

\articletype{Article Type}%

\received{<day> <Month>, <year>}
\revised{<day> <Month>, <year>}
\accepted{<day> <Month>, <year>}

\begin{document}

\title{LibreSocial: A Peer-to-Peer Framework for Online Social Networks}

\author{Kalman Graffi}

\author{Newton Masinde*}


\authormark{GRAFFI \textsc{et al}}

\address{\orgdiv{Technology of Social Networks},\\%
 \orgname{Heinrich Heine University},\\%
 \orgaddress{\state Universit{\"a}tsstrasse 1, 40225\\
  D{\"u}sseldorf}, \country{Germany}}

%

\corres{*Newton Masinde,\\
 \email{newton.masinde@hhu.de}}


\abstract[Abstract]{Distributed online social networks (DOSNs) were first proposed to solve the problem of privacy, security and scalability.
 A significant amount of research was undertaken to offer viable DOSN solutions that were capable of competing with the existing centralized OSN applications such as Facebook, LinkedIn and Instagram.
 This research led to the emergence of the use of peer-to-peer (P2P) networks as a possible solution, upon which several OSNs such as LifeSocial.KOM, Safebook, PeerSoN among others were based.
 In this paper, we define the basic requirements for an P2P OSN. We then revisit one of the first P2P-based OSNs, LifeSocial.KOM, that is now called \textit{LibreSocial}, which evolved in the past years to address the challenges of running a completely decentralized social network.
 Over the course of time, several essential new technologies have been incorporated within LibreSocial for better functionalities.
 In this paper we describe the architecture and each individual component of LibreSocial and point out how LibreSocial meets the basic requirements for a fully functional distributed OSN.}

\keywords{peer-to-peer; framework; social network}

\jnlcitation{\cname{%
  \author{K. Graffi} and \author{N. Masinde}} (\cyear{2020}), \ctitle{LibreSocial: A P2P Framework for Online Social Networks}}

\maketitle


\section{Introduction}
\label{sec:Intro}
The analysis of online social network (OSN) trends over a period of more than a decade has shown significant growth in their popularity among users, and consequently the number of OSNs has risen significantly~\cite{PZD11,GPD16}.
This growth in the users and the OSNs is directly attributable to advances in computing technologies (both hardware and software), and increased computer know-how among the users.
Current popular OSNs rely heavily on the centralized computing model, in which the OSN service provider is in charge of handling and presenting the data of the users, and actually owns all the data.
Besides the more obvious technical risks in the centralized computing model for OSNs which the providers have endeavored to address, such as unbalanced load distribution, performance bottlenecks, single point of failure, single point of attack and channel bottleneck~\cite{BBK+15}, we see two other concerns that centralized OSNs have not addressed well: accumulated costs which manifest as scalability concerns~\cite{GCR13,MKI+16} and security and privacy concerns~\cite{GCR13,AAA+15}.

The first concern, \textit{accumulated costs due to scalability}, are introduced because of a large number of highly connected users, need for more infrastructure due to a large network, high network traffic, need for mechanisms for management and dissemination of the user-generated content and challenges associated with database scalability~\cite{MKI+16}.
To cover the rising costs, service providers tend to monetize the users' content and private data, by selling it to third parties.
The second concern, \textit{security and privacy}, are divided into three categories, that is, user-related, service provider-related, and third-party application related threats~\cite{GCR13,AAA+15}.
User-related threats are a consequence of disclosure of private data to other users intentionally, such as by hacking, or unintentional due to lack of or poorly configured privacy settings.
service provider-related threats are mostly due to the fact that the service provider has control of users' data.
While the user must trust the provider to treat his data properly, the provider can leak the user's personal data outside the context of its initial definition~\cite{KWi09,ARu12} and can further allow information linkages by unauthorized third parties who aggregate data from different social data centers to obtain more information about the different users~\cite{KWi08}, such as in the Facebook-Cambridge Analytica data scandal \footnote{https://www.bbc.com/news/topics/c81zyn0888lt/facebook-cambridge-analytica-data-scandal}.
Third-party application related threats are introduced by users to provide extra functionalities that are not in the OSN, and in most cases they are untrusted.
For users to use them, they must allow the application to access their private data which then exposes them and in many cases there is no component that screens how the application manipulates the user's data~\cite{BCh09}.

Based on these concerns, researchers have proposed the use of decentralized or distributed computing models hence the emergence of distributed online social networks.
A \textit{distributed online social network} (DOSN) is described as ``an online social network implemented on a distributed information management platform, such as a network of trusted servers or a peer-to-peer (P2P) system or opportunistic network''~\cite{DBV10,CDG+14}.
It is distributed in the sense that all computing, storage and communication resources are provided by the user rather than an economically driven provider. I
This allows shifting of the implementation of the infrastructure, and the privacy and security control to the users, while allowing users to undertake innovative development of the system, effectively lowering the operational costs~\cite{BuD09}.
DOSNs can be realized via two modes of implementations, \textit{web-based} and \textit{peer-to-peer (P2P)} DOSNs~\cite{PBS11}.
Web-based DOSNs are also sometimes collectively called hybrid OSNs.
They are heavily reliant on distributed, federated web servers usually referred to as pods, that are often operated by private individuals.
Operating a pod requires in-depth technical skills, which eventually limits the features as well as concentrates the data at a few pods run by capable users. 
Inexperienced users can join existing pods. 
This renders users vulnerable to privacy concerns as the pod operators can still access private data on their pods. 
Similarly to centralized solutions, the majority of users have to trust someone to maintain the privacy of their data, a trust that is often misused.
The P2P DOSN (or P2P OSNs) on the other hand operate through the sole interconnection of P2P software run on user devices, similar to previous P2P file sharing networks, requiring no trust in any operator but require more complex networking and system solutions.
With the right combination of P2P mechanisms, such as presented in this paper, a fully distributed, scalable, reliable and secure OSN platform can be created that purely runs on the ``free'' computing, storage and networking resources of the user devices. 
By this, the P2P DOSN does not require the invest of money and thus is free of the need to monetize its service. 

Peer-to-peer networks are a class of distributed networks in which the peers act simultaneously as servers and clients, that is providing and consuming resources to/from each other, in a self-organizing manner without centralized control. 
The features of P2P networks include high degree of decentralization, self-organization, multiple administrative domains, low barrier to deployment, organic growth, resilience to faults and attacks, and abundance and diversity of resources~\cite{BYu10,RoD10}.
However, in spite of these promising features, existing P2P research offers only fragments for a working DOSN, as each individual component, such as the overlay network, data management and routing, comes with its own challenges. 
Aiello and Ruffo~\cite{ARu12} argue that (traditional) P2P systems are in and of themselves not a complete solution for a couple of reasons.
Firstly, P2P systems such as the structured overlay networks have, in their unmodified state, many security challenges~\cite{UPS11} that make the network unstable.
Also, within the P2P open environment, data access is typically open to everyone which compromises the user's data privacy.
Lastly, the most P2P overlays offer only very restricted and low-level APIs while social applications need a suite of higher-level services to reduce the overhead during the application development. 
Thus, individual p2p mechanisms discussed in literature are not ready to user for a P2P DOSN, but require carefully extension, adaptation and integration to address the needs in a P2P DOSN. 

In this paper, we present LibreSocial, that realizes a satisfactory level of reliable functionality in form of a P2P framework, adequate social interaction applications on top and meets the security, privacy and the essential quality of service (QoS) requirements for a secure decentralized OSN. This work on LibreSocial is motivated by three reasons.
Firstly, the previous works~\cite{GPM+08,GGM+10,GGS11} discussed and evaluated the security and privacy aspect of the framework, and~\cite{Gra10} focused evaluating the monitoring functionality of the system.
Therefore, although the system was introduced in parts, the finer details of the framework were not discussed.
Secondly, the framework has experienced many changes over the last five years, maturing and leading to key insights on the interdependencies of the components. 
Finally, we present the implementation and an evaluation of LibreSocial, as current most of the P2P-based OSN proposed in literature have no tangible implementation that can be tested in a live environment.
The rest of the paper is organized as follows.

In Section~\ref{sec:RelatedWork}, we look at several proposed P2P OSNs, and discuss their achievements as well as shortcomings.
In Section~\ref{sec:p2p_OSN_framework}, we introduce the concept of the P2P framework by first discussing the technical requirements needed for the framework and thereafter introduce \textit{LibreSocial} (previously called \textit{LifeSocial.KOM}~\cite{GPM+08,GMM+09,GGM+10,GGS11}).
In Section \ref{sec:p2pOverlay}, we introduce the core functions of the P2P overlay which builds on FreePastry, an implementation of the Pastry overlay~\cite{RoD01}, highlighting the modifications so as to suit our application's needs.
Section~\ref{sec:p2p_framework} to~\ref{sec:GUI} discuss the essential framework component layers of LibreSocial's system architecture, namely the overlay, the p2p framework, the OSN plugins and the GUI, showing how the defined requirements are realized.
We describe the evolution of LibreSocial from its former version LifeSocial.KOM in Section~\ref{sec:Comparison} and present in Section~\ref{sec:Evaluation} the evaluation of LibreSocial. 
A conclusion and outlook for future work is given in section~\ref{sec:Conclusion}.

\section{Related Work}
\label{sec:RelatedWork}
An analysis of majority of the DOSN solutions proposed distinguishes two main research directions~\cite{ARu12}.
The first direction focuses on the design of fully-decentralized privacy-aware OSNs.
The second direction tries to meet the privacy goals while assuming an existing central content provider.
In this paper, attention is drawn towards the provision of a fully-decentralized P2P-based OSN solution, run purely on the devices of the users, identifying some examples in literature that offer P2P-based OSN that are designed based on a clear framework or architecture.

\subsection{Peer-to-peer OSNs}
\label{subsec:p2pOSNs}
There are several proposed OSNs since the advent of LifeSocial.KOM\cite{GPM+08,GMM+09,GGM+10,GGS11} in 2008, which aim at a P2P DOSN platform run purely on the user devices, mainly to address the privacy concerns identified in centralized OSNs.
In this section, we introduce the most prominent of these proposals and also briefly mention shortfalls that we have observed in them. 



\textit{PeerSoN}\footnote{http://www.peerson.net}~\cite{BSV09} is designed with the aim of addressing privacy concerns and ensuring availability.
In the system, a solution for the privacy concern was provided by integration of encryption and access control mechanism to give a unified user login procedure.
Availability was addressed through implementation of file sharing procedures.
The architecture of the system is two-tiered in nature and is designed to decouple the user contents from the control mechanisms.
The first tier is made up of the users and their content, wherein the users can exchange their content directly.
The second tier is the DHT that provides lookup services which users utilize to locate resources, after which they interact with them directly.
This system does not provide much more than secure content-sharing as would be expected in the sense of a complete OSN. Therefore, users cannot interact with each other through messaging services.
Also, the system does not offer a replication scheme while it stores offline messages at the DHT. Additionally, the system fails in guaranteeing content privacy, user anonymity and identity management.

\textit{Safebook}'s~\cite{CMS09,CMS09b,CMO10} design is aimed at solving three key challenges, content privacy, resource availability and secure end-to-end communications.
Its infrastructure is made up of three components.
The first consists of several \textit{matryoshkas} which are social trust based topology structures that provide distributed data storage.
The second is a peer-to-peer substrate, which is essentially a DHT that provides lookup services.
The last component is a trusted identifier service (TIS) that ensures protection against DHT type attacks such as Sybil attacks~\cite{Dou02} and impersonation attacks.
The TIS does not participate in data management which guarantees privacy preservation.
Message integrity and confidentiality is possible due to the secure end-to-end communication that provides encryption and decryption via a public key infrastructure (PKI).
Like PeerSoN, Safebook also provides secure content-sharing.
In addition, it provides an identity management mechanism and also guarantees user privacy and anonymity.
The secure transfer of files require a time consuming multi-hop transmission along trust relationships in the network. 
However, Safebook requires trust in friends and will not work without friends, thus faces a bootstrap problem. 

\textit{Porkut}~\cite{NPA10} and \textit{My3}~\cite{NPA11} are similar proposals by the same authors.
They both focus on enhancing the replication scheme of the system by taking advantage of the overlaps in online times of trusted friends.
The features of the systems include: a DHT to store meta information of the user based on a user to trusted proxy set (TPS) mapping; an online time graph with the friends of the user as vertices and edges being existent if there is an overlap in online times between trusted pairs; and a storage layer that is made by constructing the TPS of a user.
In these proposal, although certain aspects of security are included such as trust, access control through the TPS and privacy preservation of the index of the profile content and meta information as well as possibility to use signatures to test authenticity of the DHT entries, they generally lack a clear definition on the features needed for a fully functional OSN. Also the systems themselves have not been converted into working applications. 

\textit{eXO}~\cite{LNT+11} is a system that is designed to offer social networking services based on a P2P platform.
The system is designed to achieve true autonomy as well as provide support for full user control while sharing content.
The public network is made up of the DHT such as Chord~\cite{SMK+01,SML+03} or Pastry~\cite{RoD01}.
The content and user profiles are stored at the DHT nodes and are both indexable and accessible through the DHT. This DHT structure allows queries to be done on the user profiles as desired.
The user profiles as well as user's content may be classed as public and hence indexed or private and not indexed.
To preserve autonomy and privacy, shared content is only stored at the user's node and is replicated to adjacent nodes in the ID space at owner's request.
The system however is very minimal in what it offers in terms of security, with guarantee for only access control and anonymity.
Other than content sharing, it appears that as an OSN it offers very limited features. 

\textit{MyZone}~\cite{MBH+11,Mah12} is a P2P-based OSN with the goal of solving availability, resiliency, routing, connectivity, security, traffic optimization and power management, which are problems common in centralized approaches.
The system architecture has two layers.
The first is the service layer that provides an infrastructure that is partitioning-resilient, dependable on top of a UDP environment and secure against malicious attacks.
This layer includes the certificate authority, rendezvous servers organized in a Chord~\cite{SMK+01,SML+03} ring network that supports searching for peers who have registered at the servers, the relay server which provides connectivity between peers that are not directly accessible to each other and a STUN (Simple Traversal of User datagram protocol (UDP) through Network address translators) server for UDP connections.
The second upper layer is the application layer that provides OSN specific features and functionalities such as higher level security policies and profile replication.
For power management control and traffic optimization, MyZone uses information pulling and versioning.
In general, this system may not be considered a pure P2P OSN as it utilizes servers for essential services.
The system may still be prone to certain security challenges that are present in centralized systems such as distributed denial-of-service attacks  against the certificate authority and the absence of anonymity.

\textit{Cachet}~\cite{NJM+12} was developed as an architecture for social networks that guarantees security and privacy while simultaneously supporting the OSN functionalities.
It utilizes a hybrid structured-unstructured overlay in which the DHT is augmented with social links formed by leveraging the social trust relationships between users.
The DHT acts as the base storage and the social links allow for the formation of a gossip-based social caching algorithm which reduces the cryptographic and communication overhead.
The social contacts act as caches that store recent updates.
The data objects are stored in decrypted form (reduces computational overhead) in containers that are protected by a cryptographic structure.
The system uses attribute-based encryption (ABE) techniques for access control. 
This system can however heavily relies on the existence of friends and trust in them, which implies that if there are no friends or trust, the OSN may not work.

\textit{DECENT}~\cite{JNM+12} is a decentralized OSN architecture that is designed to meet the need for flexible data management, support security by providing confidentiality and data integrity mechanisms as well as access control policies.
Additionally the architecture ensures data availability by utilizing relevant DHT functionalities.
DECENT has a modular architecture which separates the data objects, cryptographic mechanisms and the DHT functionalities, allowing them to interact through interfaces.
This modularity offers to the system designer the ability to make a choice of which cryptographic mechanism or DHT type to use.
However, though the system has been presented in literature, it is still unclear as to the existence of a running system that can be compared with in the public online community.

\textit{LotusNet}~\cite{ARu12} is a framework for the development of P2P-based social network services designed to support strong user authentication but with a trade-off between security, privacy and essential services within the DOSN. Its design allows for defining of privacy settings independent from the system thus users can fine-tune their own privacy configuration from a selection of several possible privacy policies.
It utilizes Likir~\cite{AMR+08} as DHT which, first, ensures two-way authentication thus mitigating against threats such as Sybil attacks, and secondly, by attaching owner signed certificates to contents it ensures secure identity-based resource retrievals.
LotusNet provides a custom suite of widgets that interact with each other in two ways, by exchanging objects through the DHT which provides the essential network services via the overlay API, as well as directly with each other if needed.
Although the proposal seems promising, it is however only a proposal and is not yet realized as a working system.

\textit{SuperNova}~\cite{ShD12} is a proposed system that relies on a super-peer network of volunteer agents and it was designed for purposes of provision of flexible storage.
The users then have a choice of where to store their content and whose content they want to store.
It also includes access control mechanisms through that allow for three levels, that is, public (visible to all), protected (visible to a selected few) and private (not visible to others), and consequently, the system also provides for full content ownership.
Data availability is enforced by replication to a list of users called the Storekeepers.
The super-peer nodes take part in the formation of the network control infrastructure, are essentially the backbone of the network and offer services such as lookup, storage, bookkeeping, recommendation and others.
The shortcomings with this OSN are: it only supports content sharing and does not provide (private) communication options, lacks essential security features such as anonymity and secure communication, and it is also does not offer an active implementation.

\textit{DiDuSoNet}~\cite{GAD+15} has the goal of utilizing the existing trust relationships to offer needed services, in particular, trustness, information diffusion and data availability.
The system is designed with two layers.
The first layer is a trust-based P2P social overlay in which the connections between the nodes are similar to social relations of the Dunbar-based ego networks of a user, with a user's friend being limited to a predefined number of approximately 150 friends called on the Dunbar number~\cite{DSh07,Dun09}.
The second level is the DHT Pastry~\cite{RoD01} which provides lookup services and ensures the system remains robust to churn.
A data availability service was implemented on the DHT. The system does not implement any security features but gave suggestions on the use of attribute-based encryption or a cipher-text attribute-based encryption for security and asymmetric keys to support access control.
The shortcoming in this proposed system is that it does not seek to implement a functional system, but rather the testing of a trust relationships and data availability via simulation.

General shortcomings in literature on P2P DOSN are thus mainly the lack of implementation and thus of a full picture of the overall system, the limitation of functionality, as well as the heavy reliance on trust in friends. 
Based on our decade long experience on building P2P DOSN, we find these three points essential to address for a suitable P2P DOSN. 
There must be an implementation attempt to identify the shortcomings and interdependencies of the considered mechanisms and to advance the architecture.
A full set of functions, namely identity management, access-controlled storage and secure communication must be provided, in order to meet the demands of an OSN and the address the technical challenges in the combination.
And finally, the solution must not assume trust in friends, as OSN use cases require to store highly personal information, often referring to these ``friends'', without them having access to it. 
In LibreSocial, we address these three challenges and more.

\section{A P2P Framework for Online Social Networks}
\label{sec:p2p_OSN_framework}
Designing P2P-based OSNs is a non-trivial task as P2P solutions usually involve some design complexities.
P2P mechanisms for realizing similar functionalities as in the centralized OSNs can be implemented in a variety of ways.
However, the necessary P2P components can be clearly defined into a suitable P2P framework, while giving system designers independence to chose the components based on their need.
In this section, we begin by underlining the necessary technical requirements in the design of the P2P-based framework to support an OSN application.
Thereafter, we introduce LibreSocial, an OSN designed based on the defined technical requirements.

\subsection{Technical Requirements for a P2P-based OSN}
\label{subsec:OSN_Requirements}
In a centralized OSN, the server stores the data, replies to queries and enforces the access control as well as other security considerations.
In P2P solutions the same must be achieved in a decentralized fashion, while ensuring quality solely based on the cooperation of unreliable and potentially malicious user devices (nodes).
First, is the need for a reliable (overlay) network that interconnects all nodes, integrates an identity management and supports routing messages to nodes/users. 
Mechanisms to store data reliably and securely with a fine-grained access control mechanism are needed. 
The provision of security and access control is especially challenging, as no one is to be trusted in the network, including friends. 
Trust in friends must not be assumed since friends in OSNs are typically known but may not be fully trusted individuals.
Beside the data-focused functions, a variety of user-to-user communication options have to be provided, allowing for secure message exchanges with dedicated recipients (unicast/multicast), such as, for chatting, gaming or video conferencing.
Once all basic technical functions are provided by a P2P framework, and a reliable and secure P2P basis of an OSN is created, the various functions of an OSN are to be implemented, such as the profile, albums, chat or news feed, and combined in a common graphical user interface. 
In the following, we structure the requirements in more detail.

\begin{enumerate}[label=(\alph*)]
\item
 \textit{Identity management}: Users should be uniquely identifiable using a set of secure, non-replicable credentials which are secured to prevents identity hijacks.
 This can be achieved through an elaborate, decentralized registration and login process utilizing cryptographic algorithms for security and privacy~\cite{GMM+09}.
 Access to an account should be purely based on knowledge, as users should be able to login on various devices and thus all necessities to log in should not be bound to a specific device.

\item
 \textit{Efficient routing}: Routing protocols are essential to build further data and  communication structures and require a reliable, authenticated message delivery.
 The routing strategy is dependent on the P2P overlay chosen, with structured overlays providing reliable ID-based routing.
 These overlays offer logarithmic routing complexity for a given ID, and the capability of defining responsibility for actions and data in the system because to the key-based routing (KBR) interfaces~\cite{DZD+03}, such as distributed hash tables (DHTs), ensuring keys are associated with addresses (for data, for nodes, for users) in a given address space.

\item
 \textit{Relevant communication channels}: A powerful and fast communication platform ensures efficient message passing.
 Mechanism that support various communication strategies such as synchronous/asynchronous, unicast/multicast or publish/subscribe should be incorporated to be able to send messages securely to individuals or groups while preventing channel clogging due to high traffic.

\item
 \textit{Content availability}: The content stored should be readily available and consistent even when the corresponding owner (or her friends) go offline. 
 This requirement is strongly dependent on the storage mechanism, the data replication mechanism and the search/indexing method that are implemented in the OSN.

\item
 \textit{Management of large data sets}: In addition to being able to store simple data within the storage layer, the system should provide structures that support the storage of large data sets such as albums or comments.
 Distributed data structures (DDSs) such as data lists (e.g. for albums), sets (e.g. friends) or on general linked data graphs (e.g. forum discussions) can be incorporated to support large data sets.
 The data structures also enable the performing of complex data queries that are normally not possible in simple key-value storage solutions.

\item
 \textit{User \&
  group management}: Users are identified through unique user identifiers (\texttt{userID}s).
 The users should be able to change their credentials without affecting their \texttt{userID}.
 The \texttt{userID} is used in friend lists, in directed communication as well as for the access permissions on the data.
 A group is a virtual user that is the union of several users and can be addressed in the communication and also in the data access permissions.
 By being able to add groups as members of other groups, hierarchies can be built that match given organization structures of the users. 
 Thus, access rights can be clearly assigned to users and groups. 
 Both users and groups can be affected by identity theft, thus it is desired that the overlay provides mitigation against such to ensure messages are sent by known and authenticated users.

\item
 \textit{Security management}: The users should be authenticated before using the network.
 The system should support the implementation of a suitable access control mechanism in combination with a suitable replication mechanisms so that users can individually pick who can access which of their data to enforce data privacy requirements. 
 The use of digital signatures to sign the content will ensure that users can verify that the data is untampered, correct and authenticated.
 Finally, the communication channels should be secured through the use of appropriate cryptographic methods for end-to-end encryption.

\item
 \textit{Quality monitoring and evaluation}: 
 Running a large-scale, distributed system always bears the risk of unpredictable behavior, this must be identified and addressed autonomously by the system itself. 
 Monitoring of the overall performance helps to expose the inherent failures and hidden strengths of the system.
 It provides a way of reviewing the quality of service (QoS) of the system.
 By reviewing data gathered from the monitoring process against the expectation on performance, strategies can be implemented that address the shortfalls in quality in specific situations.
 Continual system testing allows the developers and later the system itself to ensure that the QoS definitions are consistently met.

\item
 \textit{Customizable suite of application plugins}: It is desirable to implement the OSN application in a modular fashion.
 The various components can thus interact with each other as plugins through supplied interfaces and use the P2P functions provided by a wrapped P2P framework.
 This makes it possible to add or remove plugins without affecting the entire system, allowing addition of distributed app repositories.
 This gives the application the ability to be dynamic and more interesting.

 \item
 \textit{Appealing Graphical User Interface (GUI)}: Integrates the   visualization of various OSN plugins in a structured manner.

\end{enumerate}

In Section~\ref{subsec:LibreSocial_Intro}, the technical requirements defined here are realized in the implementation of LibreSocial, our P2P-based OSN solution, which we introduce and describe further in the consequent sections.

 \begin{figure}
 \centering \includegraphics[scale=0.225]{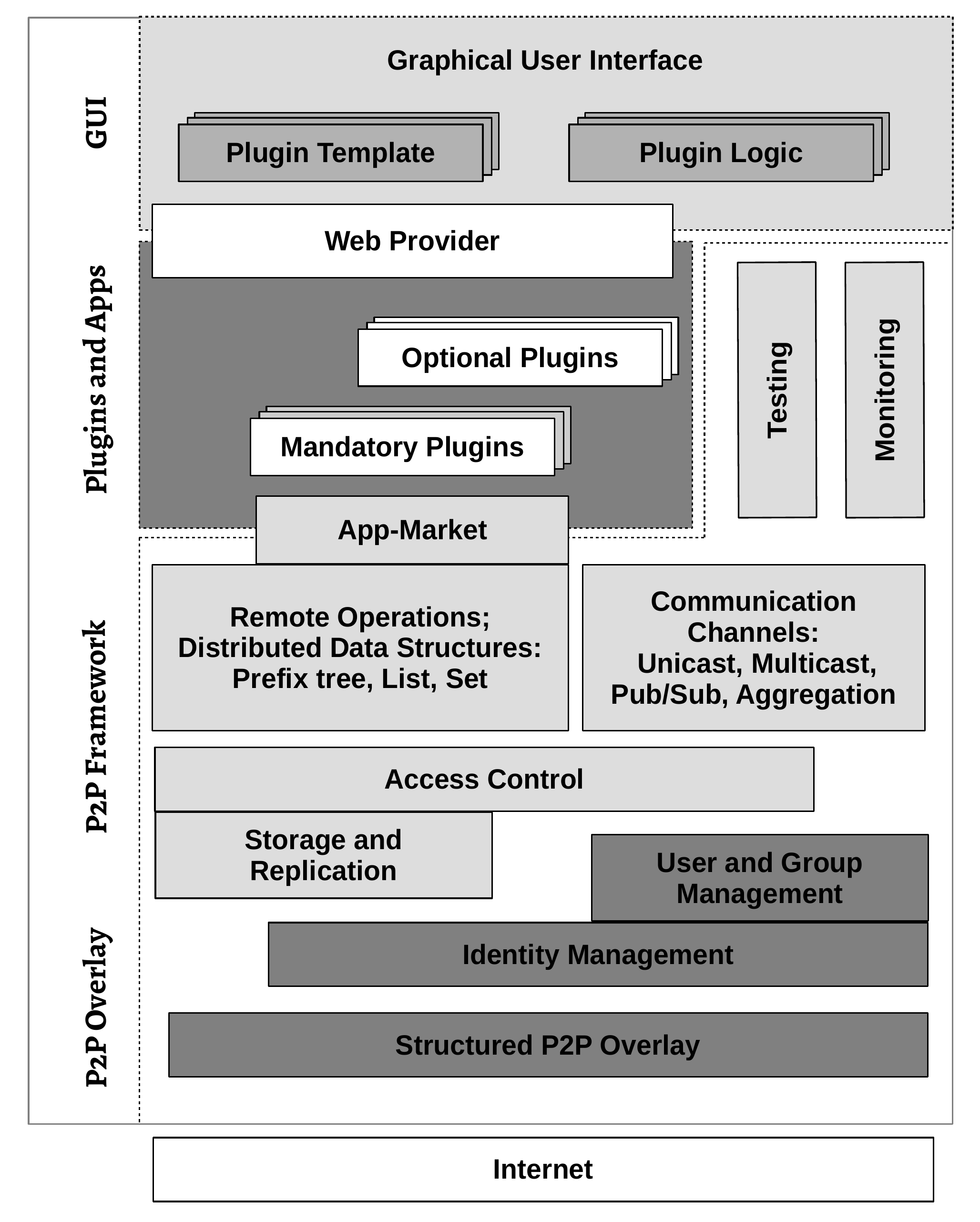}
 \caption{Overview of a proposed P2P OSN architecture}
 \label{fig:p2p_framework_overview}
\end{figure}

\subsection{LibreSocial: A P2P-based OSN}
\label{subsec:LibreSocial_Intro}
We propose LibreSocial, a P2P-based OSN that is designed based on the defined technical requirements given in Section~\ref{subsec:OSN_Requirements}.
In Fig.~\ref{fig:p2p_framework_overview}, the technical requirements are put together into a proposed framework for a P2P-based OSN. 
The architecture of LibreSocial has been designed based on the Open Services Gateway Initiative (OSGi) service platform with the goal of making it easy to add/remove services.
The architecture consists of four layers namely: 
\begin
 {enumerate*}
\item
 the \textit{P2P overlay},
\item
 \textit{P2P framework},
\item
 \textit{plugins and applications}, and
\item
 the \textit{graphical user interface (GUI)}.
\end
{enumerate*}

The most important aspect of this architecture is that it can also provide support for the deployment of any other P2P application (in form of a set of alternative plugins) by simply separating the OSN plugins from the overlay and the framework which can then be used for the new application. 
This is possible because of the strong P2P framework in the middle summarizing all essential P2P functions in an abstracted manner.
The OSGi service platform, implementing a local service bundle orchestration, further permits the adaptation of the system to any other application as desired, with possible code reuse.
Parts of this system has been previously published under the term \textit{LifeSocial.KOM}~\cite{GPM+08,GMM+09,GGM+10,GGS11} which was renamed due to naming conflicts.
In the following sections, we discuss in detail how each of the layers is realized.


\section{Overlay: A heavily modified FreePastry}
\label{sec:p2pOverlay}
Pastry~\cite{RoD01} is a generic P2P routing and object location scheme whose nodes forms a structured P2P overlay that is completely decentralized, fault-resilient, scalable and reliable.
FreePastry\footnote{http://www.freepastry.org/FreePastry/} is a readily available open source implementation of Pastry that was developed at Rice University and is extensively used within the research community.

The choice of Pastry, and in this case, FreePastry, as the overlay in LibreSocial mainly because it comes bundled with many other simple but useful P2P-based tools which are needed in LibreSocial.
These tools includes PAST~\cite{DRo01,RoD01b} (a replication scheme for simple key-value pairs), Scribe~\cite{RKD+01} (a simple multicast event notification infrastructure) and SplitStream~\cite{CDK+03a} (a multicast streaming system that uses Scribe).
These simple and highly limited tools can be directly accessed without need for further installation/configurations. 
In addition, FreePastry offers key-based routing functionality~\cite{DZD+03},  hence it strives to achieve reliable ID-based routing.
In LibreSocial, FreePastry has been heavily extended to provide secure identity management, secure and parallel routing management among other modifications to adapt it to the needs of the system.
We discuss these further below.

\subsection{Initial identity management}
\label{subsec:IdentityMgmt}
FreePastry relies on the use of DHTs for routing data in the network.
Therefore \texttt{ID} management is based on the DHT. We look at how the identity space is created and how the identifiers are constructed

\textbf{Identity space}: The DHTs utilize a predefined \texttt{ID} space of size
$2^{160}$ for all nodes which can be viewed as a circular structure in which the successor of the highest \texttt{ID} is the lowest \texttt{ID}, that is 0, hence a ring network.
Peers are responsible for the \texttt{ID}s closest to them.
Pastry defines the closest node as one having a \texttt{nodeID} with the longest possible matching prefix to the desired \texttt{ID}.
Each peer maintains a routing table with entries pointing to other peers in exponentially growing distances.
Also a \texttt{leaf set} with the numerically closest nodes in the ring is maintained. 
The construction of the routing table ensures that it is always possible to find a node that is closer to any \texttt{ID}.
If no peer is identified as being close to a given \texttt{ID}, the current peer becomes responsible for this \texttt{ID}.


%

\textbf{Identifier construction}: Every peer in the initial ring has a unique numeric identifier called the \texttt{nodeID} that is generated randomly for each node.
Each \texttt{nodeID} is a 160-bit value, with the values of the \texttt{nodeID}'s being uniformly distributed over the numeric space in which the identifiers are picked from.
This random assignment of \texttt{nodeID}'s ensures, with high probability, that nodes with adjacent \texttt{nodeID}'s are diverse in geography, ownership, jurisdiction, network attachment and so on.
The overlay also offers an efficient routing functionality.
Given a numeric value in the 160-bit numeric space and a message, the overlay is capable of efficiently routing the message to the network node whose identifier is numerically closest to the given numeric value.

\subsection{Initial Message routing}
\label{subsec:Msg_Routing}
The message routing process is made possible by the routing algorithm.
We discuss the routing algorithm and its constructs, that is the routing table and the leaf set.

\textbf{Routing algorithm}: Given a network that consists of
$N$ nodes, the overlay's routing algorithm guarantees that the message will be delivered to the recipient node in
$O(log_{2}N)$ steps.
At each routing step, the message is forwarded to a node whose \texttt{nodeID} shares key whose prefix is at least one digit longer than the prefix that the key shares with the present node's ID. If such a node is unknown, the message is forwarded to a node whose \texttt{nodeID} shares a prefix with the key as long as the current node, but is numerically closer to the key than the present node's ID. The routing algorithm in the initial FreePastry takes advantage of three data structures, the \textit{leaf set} and a \textit{routing table}, which are different for each node and help the node keep track of its immediate neighbors.


\textbf{Routing table}: This is organized into
$O(log_{2^{b}}N)$ rows
($b$ being a configuration parameter with typical value of 4) with a total of
$2^{b}-1$ entries in each row.
The
$2^{b}-1$ entries at a given row
$n$ refer to a node whose \texttt{nodeID} shares the current node's \texttt{nodeID} in the first
$n$ digits but whose
$n+1$th digit has one of the
$2^{b}-1$ possible values other than the
$n+1$th digit in the present node's ID. 

\textbf{Leaf set}: The leaf set
$L$ refers to the node set with the
$\lvert
L\rvert/2$ numerically closest larger nodeIDs, and the
$\lvert
L\rvert/2$ nodes with numerically closest smaller nodeIDs, in reference to the current node's \texttt{nodeID}.
The leaf set is especially important during the process of message routing.

Pastry further considers a \textit{neighborhood set} which holds \texttt{nodeID}s and IP addresses of the
$\lvert
M\rvert$ nodes closest (based on the proximity metric) to the current local node.
The neighborhood set is used in routing messages as well as maintaining locality properties.
In FreePastry the neighborhood set is not implemented.

\subsection{Overlay modifications}
\label{subsec:Overlay_Mods}
In order to create a foundation for security in the overlay as well as provide support for heterogeneous nodes within the context of the OSN, it is necessary to make severe changes to the FreePastry design and implementation.
These changes are summarized herein. 
We present a further in depth analysis on the construction of P2P overlays with desired properties in this dissertation \cite{Amft17}.

\subsubsection{Secure \texttt{nodeID}} In LifeSocial, asymmetric cryptography was provided using a 1024-bit RSA algorithm~\cite{RSA78} which necessitated modification on how FreePastry works so as to accommodate the 1024-bit as opposed to a 160-bit \texttt{nodeID}.
This has been changed in LibreSocial to elliptic curve cryptography (ECC)~\cite{Mill86,Kob87} with 160-bit keys which now matches the initial requirements of FreePastry while provide strong encryption with minimal overhead.
Also symmetric cryptography is provided using the advanced encryption standard (AES) algorithm~\cite{DaRi01,DaRi02} with a 128-bit key size.
We now consider the process of registration, and profile and \texttt{userID} creation, immutable and associated to a user in contrast to the \texttt{nodeID}, which is mutable and associated to a node.

\textbf{Registration procedure}: The registration process of a new user requires that there is a network, i.e. another member present to act as the bootstrap node, else the new user will become the first node in the network.
Based on the user name and passphrase the user generates an asymmetric key pair.
The 160-bit public key is used as the \texttt{nodeID} during the creation of the Pastry Node associated to this user.
The user can change the \texttt{nodeID} at will by simply regenerating a new set of keys.
With the \texttt{nodeID} being a public key, all communication to this node can be encrypted and any signatures from a node can be verified directly.

\begin{figure}
 \centering \includegraphics[scale=0.35]{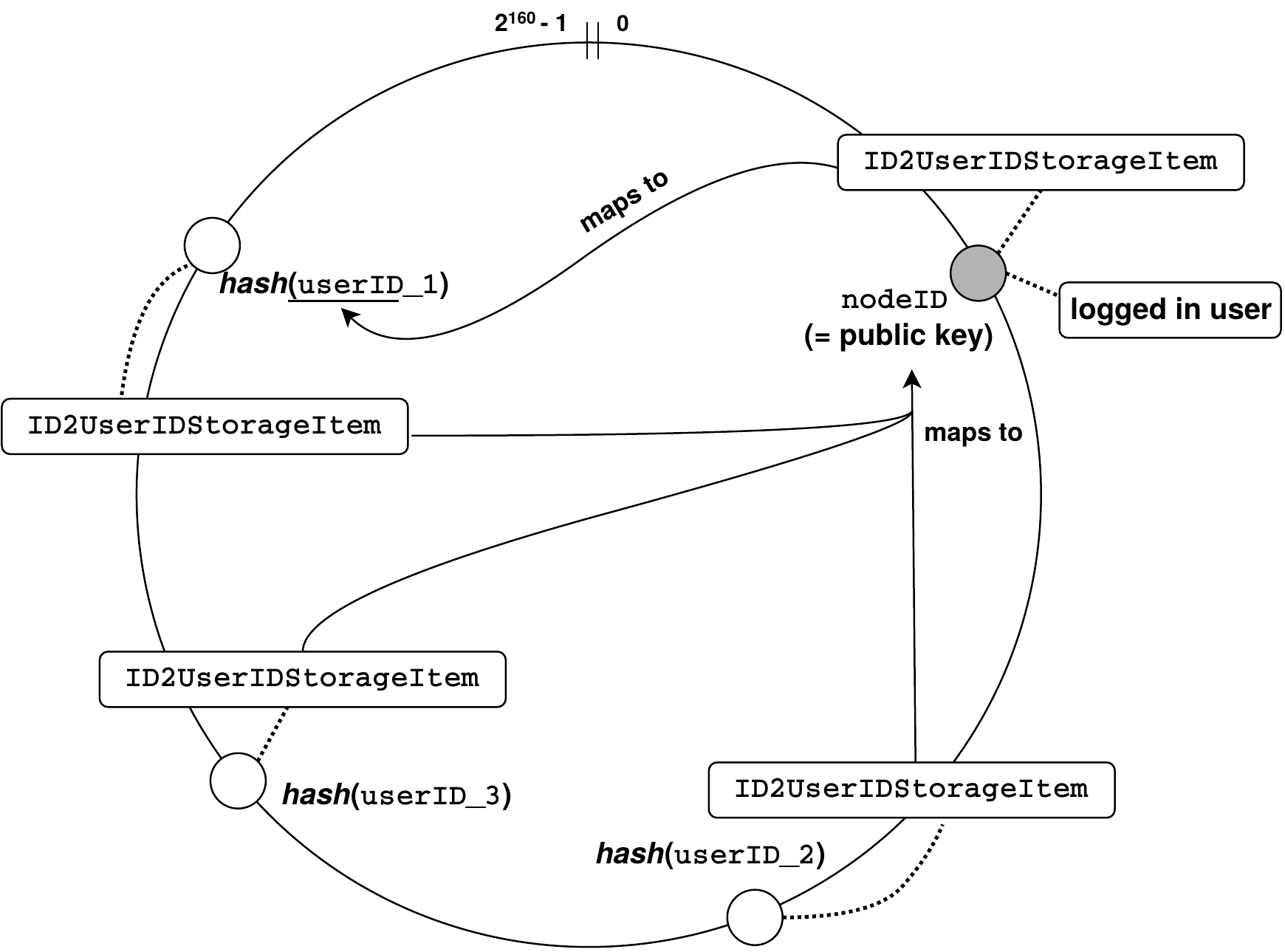}
 \caption{\texttt{UserID} to \texttt{NodeID} mapping}
 \label{fig:User2Node_Mapping}
 \vspace{-0.3cm}
\end{figure}

\textbf{Profile and \texttt{UserID} management}: A profile item is created once a new user is able to join the network, which can then be fully or partially encrypted using a symmetric key for confidentiality and stored with the public key as the \texttt{nodeID} inside the network.
Because the \texttt{nodeID} can change, it cannot be used as a unique identifier for the user, to help users identify each other.
Therefore an immutable \texttt{userID} is also generated once when the user's profile is created.
To map the \texttt{nodeID} to the \texttt{userID}, a mapping-item is created, which is stored at the \texttt{userID} in the overlay and lists what the corresponding \texttt{nodeID} to the \texttt{userID} is.
In order to prevent attackers from illegally overwriting the mapping-item, it is required that the overwriting node verifies that it is in possession of the claimed asymmetric key pair through a challenge-response approach. 
Additionally, several other mapping-items are also stored in the network to map the \texttt{userID} to the current \texttt{nodeID} so that other nodes can find out the current \texttt{nodeID} of a given known \texttt{userID}.
This is shown in Figure~\ref{fig:User2Node_Mapping}.
The routing algorithm is used to retrieve this mapping item.
By this approach, users can log in to their account from any machine, as the keys and credentials are purely created from their username and passphrase, i.e. their knowledge. 
By storing the \texttt{nodeID} (= public key) as a signed data item under the \texttt{userID}, nodes can identify the location of their contact in the network.
Any communication to this node can now be encrypted and signatures from this node verified.

\subsubsection{Parallel and iterative routing} The routing table has been extended to contain not just one entry per routing entry but a bucket of multiple peer addresses as in the DHT Kademlia~\cite{MMa02}. 
Instead of forwarding a message to only one single node from the routing table, in LibreSocial a requesting node will deliver messages to $k$ different peers in parallel.
The requesting node waits to receive the responses and then sends the message to the new
$k$ most promising peers.
This process is iterated until the target peer is identified and the messages is delivered. 
This significantly reduces the impact of interference during the message routing process by a malicious nodes and has been proven successful in Kademlia~\cite{MMa02} as well.
However, to fully realize the benefits of this technique requires two optimization problems to be solved: termination of redundant messages after successful lookup and rejection of further suggestion after forwarding of messages is done.



\subsubsection{Weak nodes} In FreePastry all participating nodes are treated as equal.
In reality, node capacities differ and some nodes are incapable of contributing to the network, either due to short participation times, missing storage capacity or limited bandwidth.
These nodes should be able to use the services in the overlay but should not be visible otherwise, that is, should not be in charge of routing and storing data.
To achieve this, the weak nodes are labeled with specific markers, implemented as specific port numbers, during the joining process.
As the IP and port numbers are available throughout the code, at any relevant position in the code it can be considered to treat weak nodes differently.
In specific, the information about weak nodes is not spread in routing, only appear in leaf sets and hence used only for final message delivery.
In the replication process these nodes are ignored.
Thus, we ensure that these nodes do not store data and do not participate in routing, both of which are taken care of by strong nodes.

\section{The P2P Framework}
\label{sec:p2p_framework}
The P2P framework is a toolbox of essential services and mechanisms that are practical and used by several applications.
The provision of these fundamental, and often required, functions is done in form of collections which smoothens the implementation of higher level applications by providing a clear interface and hiding the complexities of the underlying, challenging overlay.
The P2P framework we present provides a selected set of functions for storage and replication, access control, distributed data structures to support large data set management, secure communication, user and group management, and testing and monitoring.
It builds on the underlying P2P overlay, which itself only provides vary basic routing functions, to create a reliable and secure interface of convenient distributed functions to build either a P2P-based OSN on top or any other P2P application.

\subsection{Storage and Replication}
\label{subsec:Storage_Replication}
The importance of data availability within any OSN cannot be overstated.
While the modified FreePastry provides simple routing and a simple data storage, data can get lost if the storing node goes offline.
LibreSocial builds on PAST~\cite{DRo01,RoD01b} to handle storage.
PAST was a natural choice for handling storage because it is already integrated with FreePastry, and also includes replication management.
In the following we discuss how PAST handles files management and replication as well as the extensions to PAST to support features that were not present.

\subsubsection{File Management} When a user joins the network, the user avails a small percentage of their storage space for network usage.
This ensures that the application can perform functions such as replication.
Key file management functions include addressing and storage, and file operations which we discuss further.

\textbf{File addressing and storage}:Files stored using PAST, like nodes in the network, are also addressed using a 160-bit identifier, referred to as the \texttt{fileID}.
This \texttt{fileID} is calculated either by hashing the file itself, being then unique and unchangeable for this given file, or, alternatively, the \texttt{fileID} is created by hashing a reproducible string, such as the user name and the function of the data object, for example, under the hash of ``\texttt{Alice\_\_Albums}'' the data item listing the albums of Alice could be found.
An inserted file is stored at the node whose \texttt{nodeID} most closely matches the prefix of the \texttt{fileID}, and the node that stores this file is found using FreePastry's routing algorithm.

\textbf{File operations}: The \texttt{fileID} is used to perform \textit{INSERT}, \textit{REQUEST}, \textit{UPDATE} and \textit{DELETE} operations.
PAST was modified so as to provide the \textit{UPDATE} and \textit{DELETE} of files functionalities, which were previously not supported.

\subsubsection{Replication} The other important function that PAST provides is data redundancy support via replication which helps guarantee data availability in case the data owner is offline.
In addition, LibreSocial incorporated a data caching extension is included to improve data access.
There is also functional support for load and traffic balancing to ensure replicated data storage and access is evenly distributed among the nodes.
These are discussed herein.

\textbf{Replication process}: PAST provides replication management, so that a file stored at a node
$x$ is replicated to
$k-1$ additional nodes, where these
$k-1$ nodes are the next closest nodes based on the \texttt{fileID} and may also be found in
$x$'s leaf set.
To ensure that the
$k$ replicas of a file were actually created, every successful replicating node transmits an acknowledgment, called a \textit{store receipt}, back to the node that performed the insertion.
We adapted PAST's initial replication mechanism to not only check whether a given \texttt{fileID} is available at the replica nodes but to actually check the hash of this file, otherwise file updates would not have been propagated.
To support efficient replication, as well as reduce system and traffic overload, caching mechanisms as well as load balancing for storage and traffic are incorporated.

\textbf{Local caching mechanism}: The response time of the application is improved through the use of an internal \textit{caching mechanism} introduced on top of PAST. It holds data items recently retrieved thus reducing the need for resolving subsequent requests for the same content within the next short period of time.
The time for which the data is served from the cache is chosen carefully to limit the traffic in the network, but also to maintain a freshness of the data and to consider potential updates. 

\textbf{Storage load balancing}: There may arise a situation in which a particular node may not have sufficient storage space or may limit it for various reasons.
In such a case, a node may reject a request to store a file or replica.
However, the system takes care of such a shortfall by providing replica diversion.
\textit{Replica diversion} is used as the first option when it comes to load balancing and is aimed at balancing the load within a leaf set
$L$.
A node
$x$ that is experiencing storage space shortage and receives a request will delegate the request to another numerically close node
$y$ within its leaf set that has more storage available but was not previously selected for replicating the respective file.
If
$y$ accepts to store the data,
$x$ then stores a pointer to
$y$.
So as to guarantee availability, a pointer to
$y$ is stored in the
$k+1$th closest node
$z$, making
$z$ responsible for the replica in case
$x$ fails, thus fulfilling the need for
$k$ replicated files.
In case all the nodes in the leaf set of
$x$ have reached their storage limit, then \textit{file diversion} is initiated and the file is distributed to another part of the \texttt{nodeID} space in PAST by selecting a different salt so as to generates a different \texttt{fileID}.
File diversion ensures a balance in the remaining free storage space in different portions of the \texttt{nodeID} space in PAST and the delegation of the data storage task does not violate the anonymity of the data owner as the data is encrypted and the node storing the data cannot read it unless it has been granted permission by the data owner.


\textbf{Traffic load balancing}: In most cases, when an overload occurs, the storage capacity is not the limiting factor but the bandwidth of the node responsible for a popular item.
If an item is requested very frequently, the node might use all its bandwidth to send out that item and still not be able to process all the requests.
We extended PAST to harness, in such cases, the large group of file receivers to spread out the data themselves.
For that the responsible node maintains a list of receivers and forwards the file request to individual nodes from this list \cite{wette13}.
Which previous receiver is selected can be chosen based on various factors.
However, for popular files, many receivers are available to serve the file even further.
As this step is optional, it only improves the performance and does not induce consistency or replication issues.
In case of a file update, the list of previous file receivers is emptied and reset.

\subsection{Access control}
\label{subsec:AccessControl}
Access control mechanisms are needed to ensure that only authorized users read from and write to data items in the network.
Authorized users are generally also selected friends and these are stored in a friend list.
LibreSocial achieves access control through the use of cryptographic keys.
For write access, we differentiate between a first write operation and the following update operations.

\textbf{First write operation}: For an unused \texttt{dataID}, anyone is free to write this first instance of the data, as no access rights are violated.
The owner of the data object generates a symmetric cryptographic key which is used to encrypt the storage item.
The symmetric key is then individually encrypted with the public keys of each user or group who shall have read rights.
This encrypted data item is signed and combined with the public key of the owner as well as the list of encrypted keys builds the secure storage item, which is then stored simply in the network under the previously unused \texttt{dataID}.
The public key of the owner or group is stored together with the signed item when it is first inserted.

\textbf{File update}: It is possible to update, that is, overwrite, the data with a new secure storage item if it is signed with the corresponding private key belonging to a specified public key, after it has been verified by the node where the data object is stored.
Fig.~\ref{fig:Secure_Storage_Item} shows an example of how LibreSocial secures an item using the symmetric and asymmetric keys as well as how it provides access to the item via the list of individual encryptions of the symmetric key corresponding to the data item.
The new secure storage item is only accepted and replicated when the same data owner provides the data update.

\textbf{Read access}: Reading of the data item requires retrieving of the data and having the right private key for the decryption of the individually encrypted symmetric key which is then used to decrypt the data.
In a hierarchical group, the symmetric key is retrieved using a depth-first traversal through the group hierarchy.
Thus, anyone may have the Secure Storage Item, e.g. as replicating node, but only the users selected by the owner can decrypt and read it.



\subsection{Distributed Data Structures}
\label{subsec:DDSs}
PAST~\cite{DRo01,RoD01b} uses a DHT data structure which supports storing of single item objects.
Therefore, it is only possible to perform simple key-value lookups.
For an OSN, the goal is to be able to store large data sets, such as albums, comment lists and friend lists, and provide mechanisms that can support execution of complex search queries, such as range queries, aggregation queries,
$k$ nearest neighbors
($k$-NN) queries, multi-attribute queries and so on.
Using PAST solely as the storage mechanism will not achieve this.
Additionally, several security control measures such as access control mechanisms that support \textit{read}/\textit{write}/\textit{append}/\textit{delete} privilege checking for files should be easily integratable.

To achieve this, we build distributed data structures (DDSs) on top of the PAST's DHT. 
The DDSs consist of a \texttt{dataID}, usually a unique hash function under which they are referenced, an optional payload, as well as an optional list of pointers to other \texttt{dataID}s~\cite{GPM+08}, allowing to build data graphs.
The object is retrievable from the DHT if the \texttt{dataID} is known.
This offers two possibilities of designing the DDS. In the first possibility, the item is stored under a hard-coded hash which is known to all network nodes.
The items in this case are hard-coded entry points for the structure traversal.
The second possibility is to store the object under its own hash and then store the pointer to this hash in another item.

To support random data retrievals within the network's graph structure, the first design option is preferred, that is, the item is stored as an entry point.
For example the album set of user \textit{Alice} is stored under the hash of the string \texttt{"Alice-Albums"}.
In this situation, due to many abstraction layers and the high level of redirection that occurs, high latency may occur during the process of the pointer/hash initiated traversals in the P2P overlay.
In any case, every level of redirection, including multiple hops, eventually leads to a 1-to-1 communication over the P2P overlay.
This scenario has the advantage that it is now possible to split some seemingly large objects, such as photo albums, into smaller parts and identify only a single object that holds a list of hashes for all the parts are pointers to further parts.
It then becomes possible to handle persistent data in a better manner, as well as provide a cleaner abstraction over the P2P overlay for the storage and retrieval of graphs connected through these hash pointers but with the possibility of latency due to redirections and additional layers.

LibreSocial supports storage of large data sets by integrating three distributed data structures (DDSs) into the framework, namely, distributed linked list, distributed set and prefix hash tree.
The DDSs are built on top of the DHT that the modified PAST provides, with different APIs and entry points for each DDS, hence no interaction takes place between them.
A discussion on the three structures follows.

\subsubsection{Distributed sets \&  linked list} A \textit{set} is a data structure that is an unordered collection of unique members/elements, and each member of the set may be either a set or a primitive element referred to as an atom.
A \textit{list} is a sequence of zero or more elements of a given type, in which the elements are linearly ordered according to their position on the list.
A \textit{linked list} is a list in which elements in the list have pointers to the next element in the list.
Sets can be represented by linked lists, where the items in the list are elements of the set.
This way the system developer can design the system without worrying about the usage of contiguous memory for storing a list, reducing the process of shifting elements to make room for new elements or closing up of any gaps created due to deletion of elements.
But this is at the cost of additional storage space for the pointers.
In LibreSocial, sets are used to manage friends, uploaded files and albums while lists are used in comment sequences, forums and in the inbox.

\begin{figure*}[!tbp]
  \centering
  \begin{minipage}[b]{0.22\textwidth}
        \centering
        \includegraphics[scale=0.18]{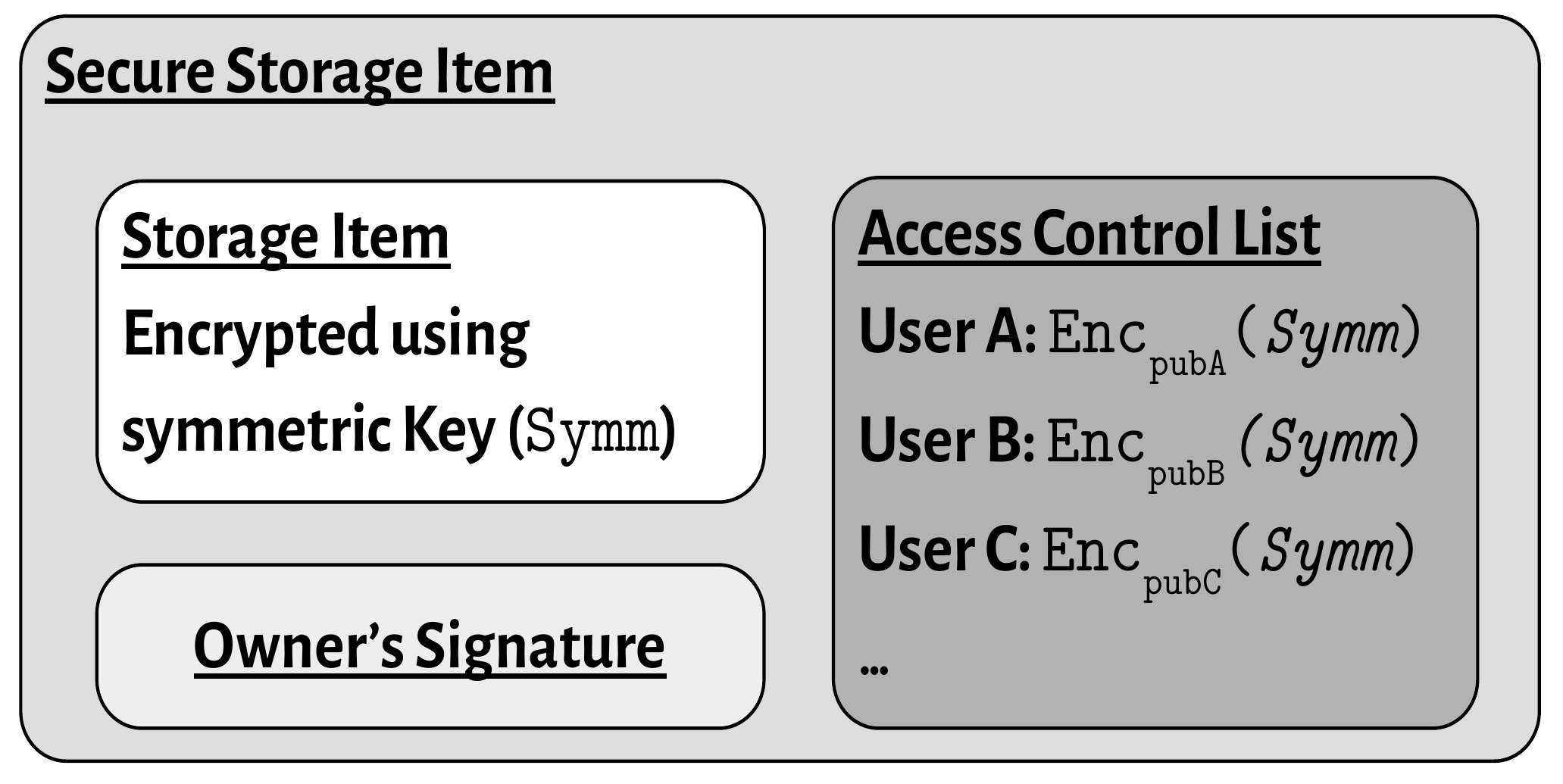}
        \caption{Secure Storage Item.}
        \label{fig:Secure_Storage_Item}
  \end{minipage}
  \hfill
  \begin{minipage}[b]{0.36\textwidth}
        \centering    
        \includegraphics[scale=0.30]{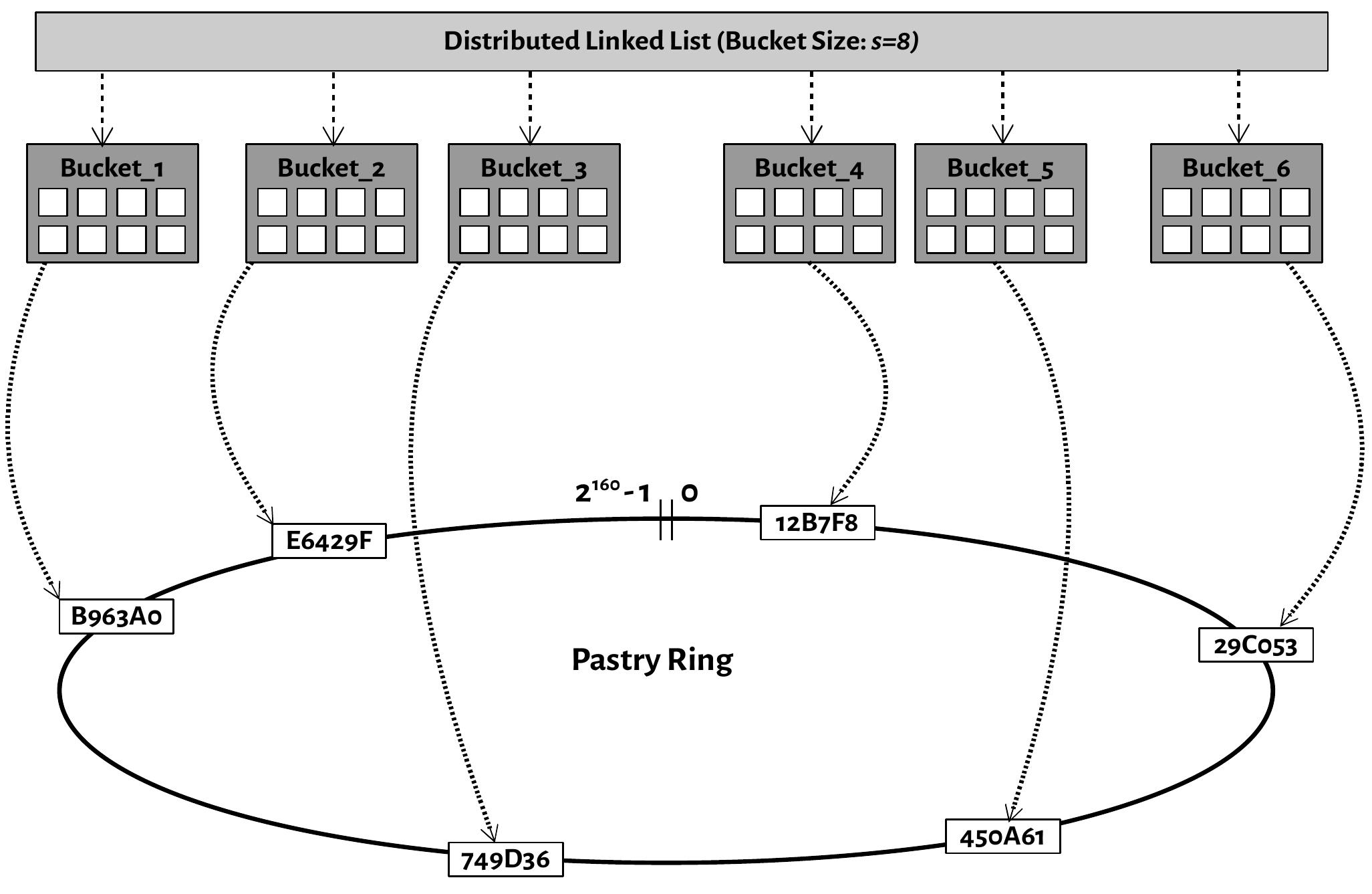}
        \caption{Structure of a Distributed Linked List}
        \label{fig:DDS_LinkedList_Mapping}
  \end{minipage}
  \hfill
  \begin{minipage}[b]{0.37\textwidth}
        \centering    
        \includegraphics[scale=0.30]{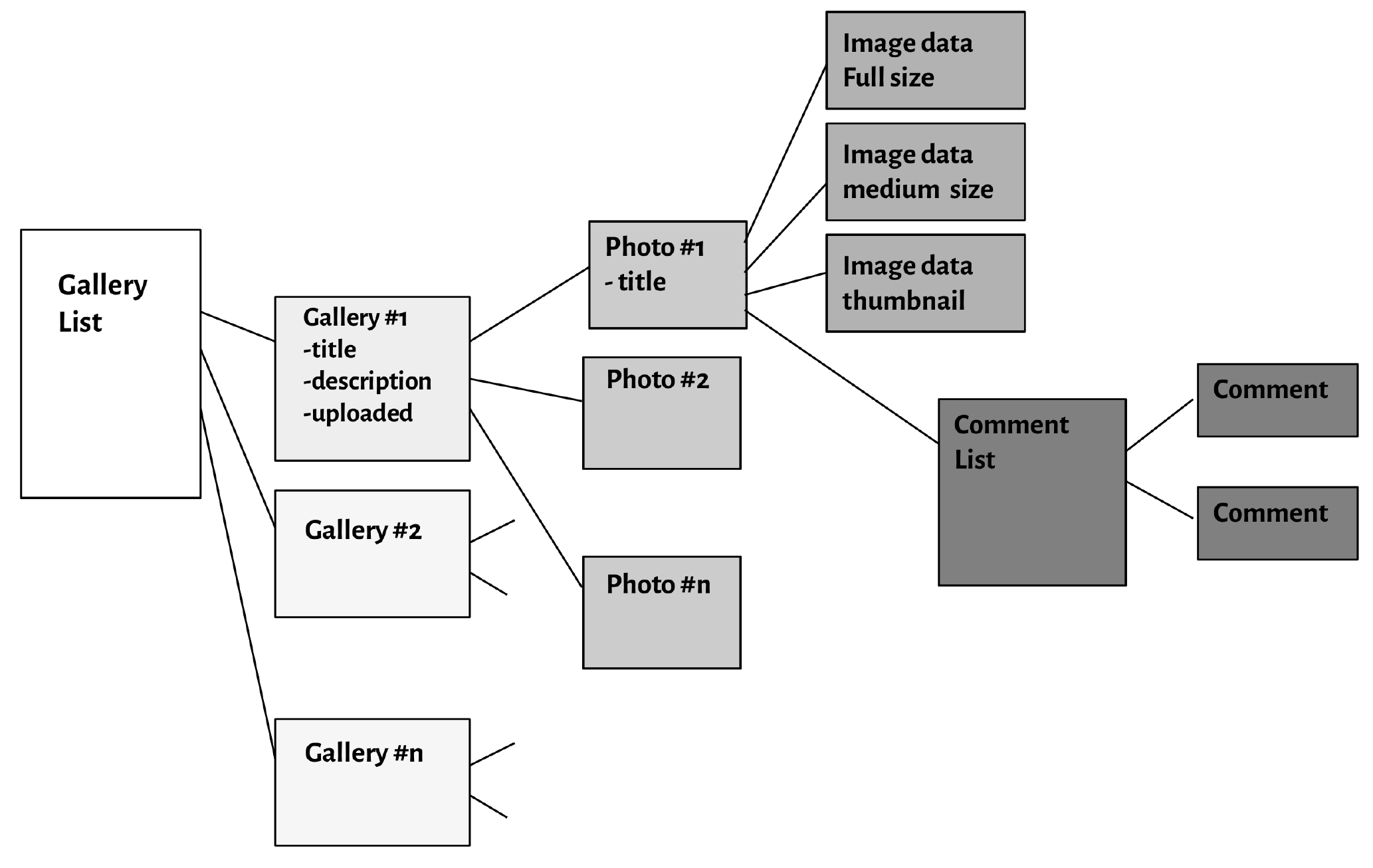}
        \caption{Photo Album using a Distributed Linked List}
        \label{fig:Example_DDS_List}
  \end{minipage}
\end{figure*}

Storage in distributed manner on various nodes is done by splitting the list of data items into buckets of a defined size, each containing a given range of the data items of the whole data structure and then distributed to the nodes in the network.
The bucket size,
$s$, depends on the size of items to be stored by the user, and is a data structure parameter.
The structure itself is in essence an array and all items from the interval
$[s\cdot(m-1),s\cdot m-1]$ for
$m\in
\mathbb{N}$ are stored in a bucket named
\texttt{structurename}\_$m$, where
$m$ is the interval ID and
$\mathbb{N}$ is a set of all network nodes.
This is represented in Figure~\ref{fig:DDS_LinkedList_Mapping}.
In case the index of an item is known, the remote node responsible for the bucket is also known and can be contacted directly, thus additional latency is in the form of one redirection.
The distributed sets/lists can be used to store structured data in form of data graph stored in the network as shown in Fig.~\ref{fig:Example_DDS_List}.

%
%

\subsubsection{Prefix Hash Trees} The prefix hash tree~\cite{RRH+04} (PHT) is a trie-based distributed data structure that was designed with the sole purpose of supporting more sophisticated queries over a DHT, specifically, it allows running of range queries, heap queries, proximity queries and multidimensional analogues of these queries.
It relies on the lookup interface of the DHT to construct a trie-based structure and does not need to know how the DHT topology looks like or how the DHT performs routing.
In LibreSocial, the use of the PHTs make it possible to perform range searches.
This allows searching for content that falls within a predefined range, such as users by age.
PHTs store key-value pairs.

Assuming an alphabet that consists only of binary digits, hence keys are represented as binary numbers, and that every leaf node stores at most
$M$ keys.
The PHT structure begins with a single node stored at a given leaf node.
Insertion of a key,
$K$, will first result in a lookup operation to locate the leaf node that stores
$K$.
The insertion process may then cause the leaf node to split into two children followed by redistribution of the keys.
Generally, the
$M+1$ keys are then distributed between the two children such that each of them stores at most
$M$ keys.
This splitting process can cascade further downwards the PHT. The PHT structure formed is a binary trie.
The beauty of the PHT is the way in which the logical trie is distributed among the peers in the network (how the PHT vertices are distributed to the DHT nodes).
This is achieved by hashing the prefix labels of PHT nodes over the DHT identifier space.
Using this hash-based assignment, it is possible to locate the corresponding PHT nodes using only a single DHT lookup.
This offers a faster method of locating data over the distributed sets and distributed linked list.
The PHT allows the range queries to locate objects.
Given two keys
$A$ and
$B$ ($A\leq
B$), the range query returns all the keys
$K$ contained in the PHT that satisfy
$A\leq K\leq
B$.

\subsubsection{DDS security and access control} All of these distributed data structures incorporate cryptographically enforced access control mechanisms.
Confidentiality is typically provided through directed encryption with the public keys of the read enabled users.
Integrity is provided through the signature of the author, which is stored alongside the encrypted/unencrypted data.
Data updates, that is, the overwriting or deletion of data, requires the signature of the previous author to take effect.
All replicating nodes consider this and thus a majority of the replica holders must confirm that the changes are valid.

The need for such advanced access control options is seen in the functionality of the news wall of every user.
This social networking specific feature gives each user a personal wall page that they can post entries to.
Other users can also post to another user's personal page and also comment on those posts if they have been accepted as friends.
Users may also alter their comments or even delete them. Entries of others must not be altered. 
Even the owner of the news wall cannot alter the entries of others, but may delete the entry in total. 
The wall is implemented as distributed list, in which the users and his friends can add entries (to the DDS list), others do not have read rights.
Also each DDS list entry is secured and may only be altered by its initial author.
A friend who wrote a comment can edit his comment by replacing the comment object under the same \texttt{dataID} with a valid signature of the same owner as the previous comment.
He even may mark the comment as deleted, that is, overwrite it with a \textit{tombstone} object, which is treated in the presentation differently.
The owner of a wall can delete a complete entry, by removing the pointer to that comment from his wall item, which he owns, but not the wording in the entry, which would required to replace the comment at the comment's data item ID in the DHT, which is not allowed, as he cannot present a valid signature from the previous owner.
The same applies for comments on photos in photo albums and entries in the forum of a group.
As the access control is enforced by the P2P framework, the applications and plugins do not have to hassle with the complexity of the access control enforcement.
In the dissertation \cite{Alaaridhi17}, we present the DDS concept in full detail.

\subsection{Communication Channels}
\label{subsec:Comm_Channels}
The communication channel refers to the point where the messages are exchanged between users.
The channel either pushes messages (which requires a callback function) or pulls messages from (which requires constant probing) from other nodes.
A message sent in a particular channel can only be received at nodes that are in this channel.
The following are aspects of the communication channel that have been designed and implemented in LibreSocial.

\subsubsection{Unicast, multicast and aggregation} LibreSocial supports both \textit{synchronous} and \textit{asynchronous} messaging.
It also provides support for \textit{unicast} (1-to-1) messaging such as in direct messaging, \textit{multicast} (1-to-N) messaging such as streaming to a group of users, and \textit{aggregation} (N-to-M) to distribute and aggregate information in the network.
P2P overlay networks support lookups which reach the destination node within several \textit{hops}.
However when there is low latency or high throughput, it is preferable to communicate directly with other nodes.
This is achieved through \textit{IP-based} communication, where the node retrieves the IP information of the other node once and uses it to communicate directly without additional hops to other nodes, such as in streaming and file transfers.
Asynchronous messaging is achieved through the storage of messages encrypted for and stored in the inbox of the recipient.
An inbox is a data structure located in the DHT, which permits the INSERT of messages, but prohibits READ, UPDATE and DELETE access to unauthorized users.
Once the recipient is online, he checks the presence of messages in his inbox, retrieves them and deletes them from the inbox.

\subsubsection{Publish/Subscribe (pub/sub)} This allows users to communicate with each other without the knowledge of each others addresses.
The messages are published to topic channels that the users have subscribed to.
Pub/sub is implemented using Scribe~\cite{RKD+01} with support for caching of created topics. 
Scribe creates a minimal spanning tree for all participants in the pub/sub channel and delivers messages from any origin to all subscribers. 

\subsubsection{Streaming via SplitStream} Streaming can be visualized as a high bandwidth 1-to-N communication, hence is different from a basic file transfer.
LibreSocial supports streaming using SplitStream~\cite{CDK+03a} which allows minimization of the upload bandwidth requirement by equally distributing the workload and ensuring that nodes which consume also participate in routing.
Support for order preservation is integrated and can be enabled or disabled depending on the application plugin requirements as this includes an overhead.
In essence the P2P framework supports transmission of low overhead streaming data that is already properly formatted.
Also, there are additional options for resending of lost packets and using checksums to verify received data.

\subsubsection{Streaming via WebRTC} While the streaming option through SplitStream routes data flows from one P2P framework instance (node) to another, there is often the need for simple, low latency audio and video conferencing.
This use case is addressed through WebRTC which is provided by the browser and allows browsers to connect to each other.
Having access to the webcam and microphone allows to set up conferencing tools.
In this case, the data flow takes place between browser instances and does not pass through the P2P framework.
This side-channel is the only exception to the aim to manage all data and communication in the P2P framework.

\subsubsection{Secure message channel} The \texttt{nodeID}s in the network are public keys.
Thus, any communication can be encrypted and signed.
To ensure that the messages are secured, they can be encrypted either by a symmetric key or using the asymmetric key depending on the communication channel that is to be used.
If the message is intended for only one recipient, then only the asymmetric key encryption is applied.
If it is meant for a group, then the symmetric key is applied and then the public key of the receiving user which is identical to the \texttt{nodeID} to which the message is sent to.
As the sender's \texttt{nodeID} is also it's public key, the recipient can easily verify the message.

\subsection{Monitoring and Testing}
\label{subsec:Monitoring_Testing}

\begin{figure*}[!tbp]
  \centering
  \begin{minipage}[b]{0.33\textwidth}
        \centering
        \includegraphics [scale=0.33]{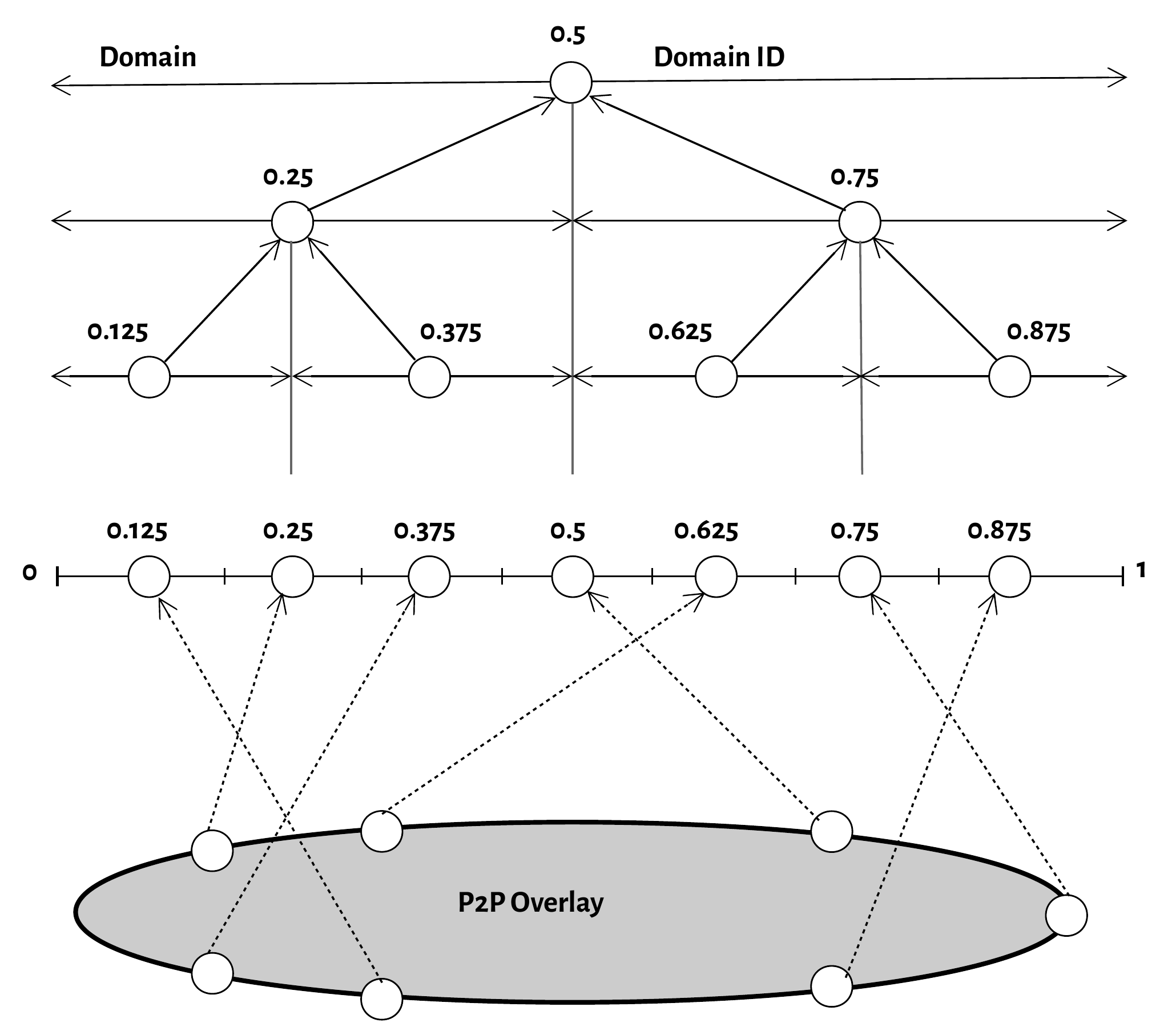}
        \caption{Tree-based Monitoring Topology}
        \label{fig:Monitoring_SkyEye}
  \end{minipage}
  \hfill
  \begin{minipage}[b]{0.30\textwidth}
        \centering    
        \includegraphics [scale=0.28]{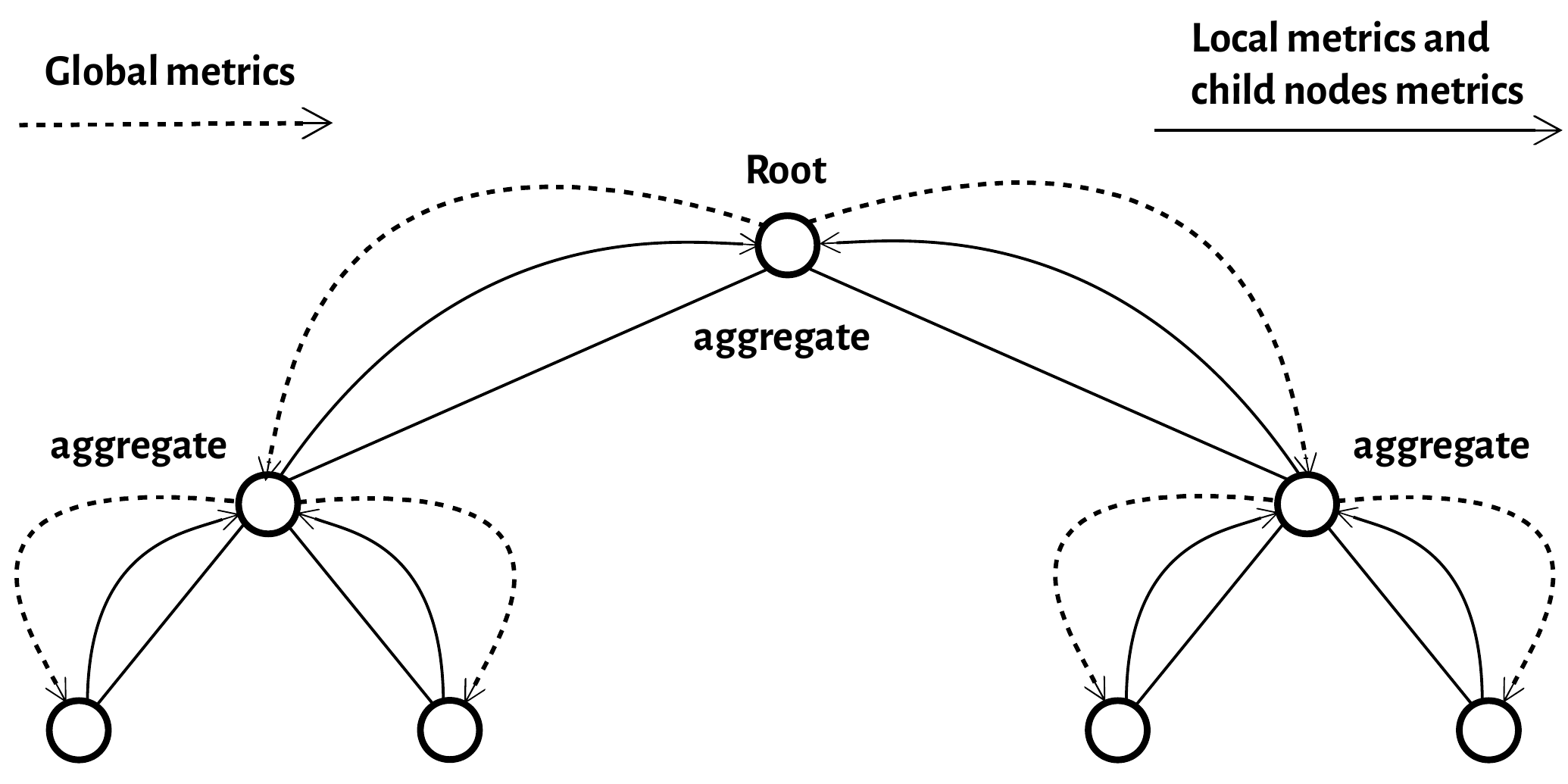}
        \caption{Metric Aggregation Strategy}
        \label{fig:Monitoring_SkyEye_Aggregation}
  \end{minipage}
  \hfill
  \begin{minipage}[b]{0.28\textwidth}
        \centering    
        \includegraphics[scale=0.25]{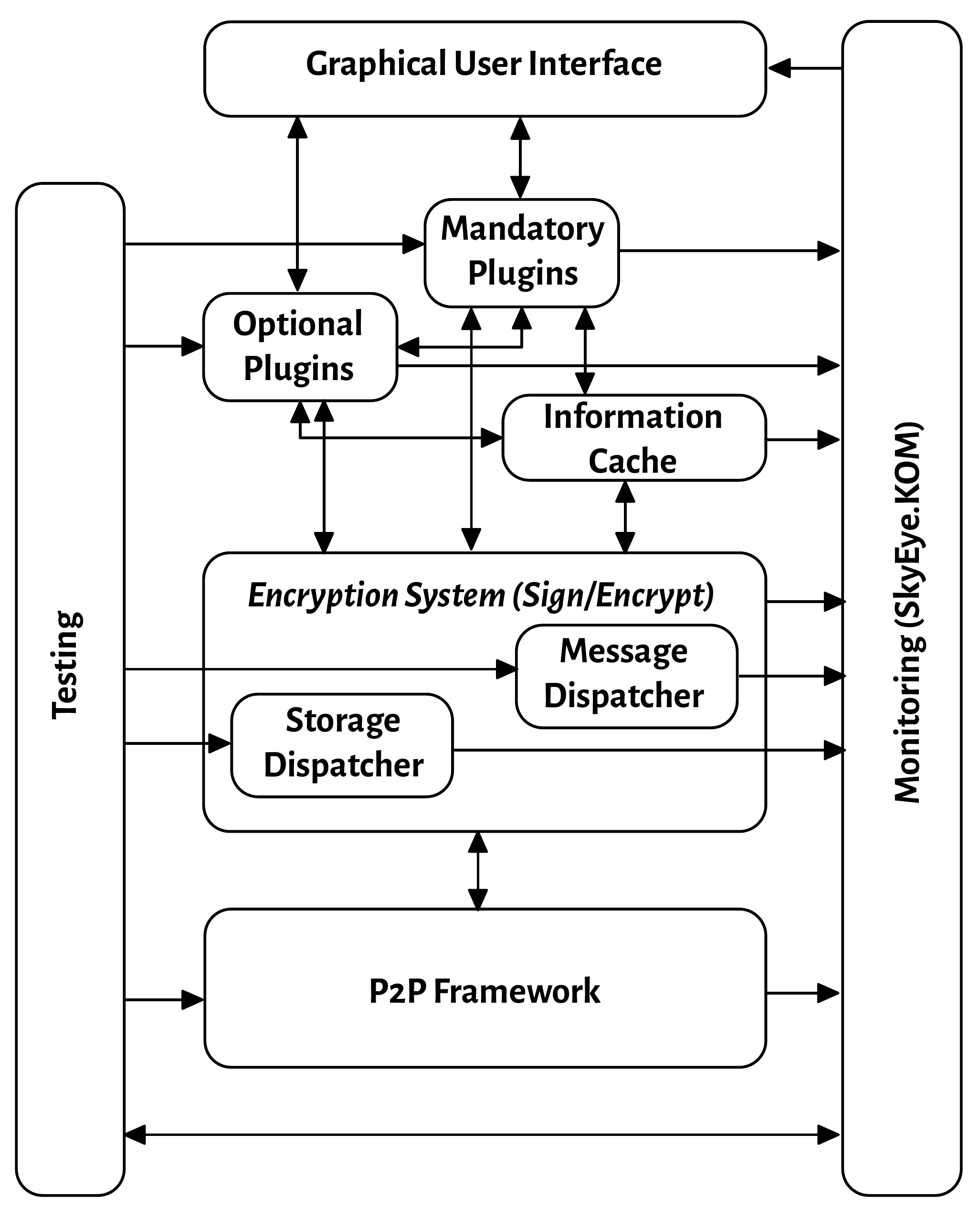}
        \caption{Component-based Framework}
        \label{fig:component_based_framework}
  \end{minipage}
\end{figure*}

Running a large-scale distributed P2P system bears the risk of unforeseen emerging performance issues.
In order to identify the quality and performance of the network, there is need for a reliable monitoring solution.
LibreSocial performs tree-based monitoring using \textit{SkyEye.KOM}~\cite{GAn17, GKX08} (or simply \textit{SkyEye}).
SkyEye works by  implementing a further overlay on top of the P2P network and arranges the nodes into a tree structure as shown in Fig.~\ref{fig:Monitoring_SkyEye}.
Every node collects local measurement for certain pre-defined metrics while at the same time receiving metrics from its own child nodes (if it has any child nodes) and aggregates these received values with its own.
After performing the collection and aggregation, the nodes sends these metrics to its pre-calculated parent node in the tree which continues this process to the root node. The parent sends an acknowledgment message (ACK). 
Messages consist of statistics, such as average, minimum, maximum, sum and standard deviation, to various metrics, such as the observed response time, bandwidth usage and much more.
Thus from the leaves up to the root of the tree, the statistics of the various measurements are aggregated and the root compiles the final aggregation: the global view on the whole network.
This information is gradually tickled down to all nodes in the network through the acknowledgment (ACK) messages.
For that, every time a parent node receives an aggregation update message from its child node, it sends as confirmation an ACK message which contains the aggregated global view that the parent received prior from its own parent node.
Thus, eventually all nodes receive the global view through ACK messages in the network.
This is shown in Fig.~\ref{fig:Monitoring_SkyEye_Aggregation}.

We elaborate the long term vision for \textit{monitoring and management} in \cite{KGr13,GSR+09} and the dissertations \cite{Gra10, Dister18}.
We describe a control loop for the P2P system to monitor itself, analyze the observation, plan a correction strategy and execute it through the distributed adaptation of the system parameters.
With SkyEye we have a view on the global performance of the network in each node.
A quorum of nodes around the root in the monitoring tree is set in charge to analyze these observation in comparison to given performance goals, such as a maximum response time or maximum hop count in routing.
In case they observe a quality degradation, counter commands are initiated, for example, to increase the routing table sizes, through the back channel to all nodes using the ACKs.
By adapting the routing table size more/less new contacts have to be established leading to shorter/longer routing paths and thus response times.
We have shown in simulation that metric intervals can be defined and reached through such a distributed control loop from both sides, that is, both in lowering the average hop count (when the response time is ``too high'') and in raising the hop count (when the response time is high but the maintenance traffic should be lowered).
However, for an integration in LibreSocial it is needed to adapt the framework so that the configuration of the system can be adapted during runtime, which is not trivial.


\textit{Testing of the system} allows the developers to find bugs that exist in the system, while monitoring allows the system users to find out exactly how the system operates in general. 
This does not interfere with the security goal, as only statistics are obtained but no personal data. 
In LibreSocial, there are two parts that are considered important for testing, 
\begin
 {enumerate*}[label=\textit{\alph*})]
\item
 the application and its plugins, and
\item
 the modifications in the overlay and the framework.
\end
{enumerate*} LibreSocial includes a TestPlugin which supports defining single test cases as well as longer test plans that execute multiple actions sequentially in predefined time slots.
Actions refer to all possible interaction options that a user can to, thus allowing to emulate user behavior in a controlled environment.
It also provides the option to randomly execute test cases. 
Every predefined test case must stipulate its own preconditions., such as having an album before uploading photos. 
The test plugin checks the fulfillment of all preconditions.
The test plans make it possible to efficiently test new plugins, updates of them as well as the performance and costs of the overlay and P2P framework in general. 
To test the various plugins, the test plugin accesses the API of the plugins used by graphical user interface which guarantees a functional API as well.

\subsection{AppStore - Repositories for (OSN) Plugins}
\label{subsec:AppMarket}
The AppStore is an independent plugin extension for repository based app/plugin management.
The development of the AppStore was possible due to the extendable nature of LibreSocial as a result of the plugin-based, OSGi architecture.
The AppStore serves as a springboard upon which users can create an exclusive \textquotedblleft AppStore\textquotedblright~that they themselves manage, similarly to the apt-repositories in Gnu/Linux.
Anyone will be able act as a publisher, host his own App repository and advertise it to various groups or all users.
Publishers can instantiate repositories, browse and search through their catalog, publish, manage, download and install different apps, as well as share the repository address (dataID) with other users. 
Users can browse the catalog and only \textit{download and install} the Apps.
A repository is shared with other users by sharing the repository ID via email, chat or direct message with them, after which they simply search for the repository ID. 
The repository is technically a DDS with metadata and fileID entries.  
The Apps are technically OSGi bundles that include the app metadata and a .jar file that can be installed dynamically at runtime by the users.
Apps are stored under a dataID using the storage functions.  
The AppStore provides three views.
\begin
 {enumerate}[label=(\alph*)]
\item
 \textbf{Publisher view}: A view to manage the content of a repository.
\item
 \textbf{User view}: Browse and install Apps from a given repository. 
\item
 \textbf{Installed Apps view}: Personalized repository listing the Apps installed by the user. 
 Allows to automatically download and install the preferred Apps when using a new device. 
 Users can \textit{delete} the Apps installed from here.
\end
{enumerate} With this setup the framework is open to any kind of further extension independent of the initial App provider of the framework, anyone can provide OSGi bundles as (OSN) Apps. 
This general applicability motivated the  P2P framework, which aims to support generic use cases and functions and does not focus on specific applications such as OSN.

\subsection{Other supporting components} In addition to the P2P features discussed in the framework, there are three other important components integrated into the framework.
These are the \textit{Storage Dispatcher}, the \textit{Message Dispatcher} and the \textit{Information Cache}.
The component framework shown in Figure~\ref{fig:component_based_framework} shows the general placement of these components in the architecture.

The \textit{Storage Dispatcher} provides storage services for the platform-specific data objects, both locally and remotely.
It keeps track of the application data being stored.
All data objects and messages in LibreSocial extend a common class called \textit{SharedItem}, having a storage key, header and the data object itself, making it a storable object.
The Storage Dispatcher acts as a local stub and performs efficient storage, retrieval, update and removal operations on the data object using the P2P framework.
Execution of these operation can be either \textit{synchronous} or \textit{asynchronous} and are managed by the Storage Dispatcher through calling the right P2P functions in the framework. 

The \textit{Message Dispatcher} is used to create instances of different types of communication channels, namely, the \textit{MessageChannel}, \textit{TopicChannel} and \textit{AggregationChannel}.
The MessageChannel interface supports bi-directional 1-to-1 communication.
It sends messages to a defined address and allows registering of a listener that is notified whenever a new message arrives at that particular MessageChannel.
Each MessageChannel has a unique name to identify it.
The MessageChannel also supports 1-to-N communication after defining a list of receivers that a particular message is to be sent to.
The TopicChannel interface uses a publish/subscribe mechanism.
It sends a message to every node that has registered to a particular TopicChannel, e.g. participants in a chat room.
The AggregationChannel is used to combine data on a global/network scale, used in SkyEye. 
In order for nodes to provide data, they must add a \textit{sensor} and to receive combined/aggregated data, the node must register a \textit{callback}.
The aggregation of the \textit{sensor data} takes place in the AggregationServer which is in the backend.

The \textit{Information Cache} acts as a cache for objects, either data objects or stored messages, requested by the higher layers from the distributed storage.
These requested objects are expected to change infrequently, hence can be kept in the cache so that subsequent requests can be served locally direct to the Apps/GUI to minimize network traffic due to repeated requests.
The cache size is configurable and the caching strategy employed in LibreSocial is the least recently used (LRU) strategy.
The use of the Information Cache allows exempting the plugins in the upper layers from handling asynchronous events.
The plugins simply decide the objects they need at any particular time and retrieve it from the cache.
Such data may either be available, already requested or not available.
In case of unavailability, the plugin then requests for it, leaving the cache to initiate the lookup for the requested object and to process the irregularly incoming data.

\subsection{Summary of the P2P Framework} The framework is a collection of (advanced) P2P functions to harness the resources in the overlay, hide the complexities and to provide interfaces for advanced applications on top.
PAST provides replication functionality and has been extended to support access control, security, heterogeneity and also load balancing.
With the mechanisms for distributed data structures and advanced communication options which all consider the security of the users various applications can be built.
The monitoring solution SkyEye supports the network by gathering and providing continuously information on the network's performance that can be used to fine tune the nodes' configuration in the system.

\section{The Plugins and Applications}
\label{sec:Plugins+Apps}
Plugins are software components that add a specific feature to a system to enhance the system's capabilities.
The use of plugins in system design provides for increased extensibility, simplicity in system design and parallel development of a software application.
The plugins used in LibreSocial are based on the OSGi framework and are placed on top of the P2P framework layer and rely on the services that it provides.
Each plugin also provides an \textit{OSGi command} interface that allows the Test plugin to check functionality of the plugin during distributed tests.
The following are the plugins that were implemented.

\begin{itemize}
\item
 \textit{Login}: This is the entry point into the network.
 The plugin handles the user's registration and login in the underlying framework.
\item
 \textit{Profile}: This plugin is used to create a data item that contains the user's personal data.
 The item is stored in the secure P2P storage.
 It allows for granular adjustment of the private data that users can view.
 They can also set cover and profile pictures.
\item
 \textit{Notifications}: Informs the user about new events within the network such as a friendship request or an invitation to chat.
 These events are stored in a distributed list as encrypted notification objects so that they can be retrieved later by a user who was offline when the event occurred.
 Other users can insert these encrypted messages in the inbox of the designated offline user, viewing or modifying existing ones is not possible. 
\item
 \textit{Files}: 
 Allows uploading, downloading, deleting and sharing of files with other users. 
 When a user selects a file to upload, this file is fragmented into chunks and stored (and replicated) as distributed data structure through the P2P framework.
 After the upload of all chunks, an additional storage item that contains all the IDs of the file chunks is generated and stored.
 Links to this file can be sent and used in LibreSocial. 
 \item
 \textit{Search}: Gives the user the ability to search for users registered in the network based on searchable (opt-in) information in their profile such as name, city, country and gender.
 \item
 \textit{Friends}: Manages the relationships that exist between the users and maintains the userIDs of friends in a DDS Set.
 It supports sending of friendship requests to other users, adding, declining requests as well as blocking and unblocking specific users from sending friendship requests.
 Also, it manages the login statuses set by the users.
Friend lists can be turned public or private. 
\item
 \textit{Messaging}: This is comparable to an email application.
 It allows users to send messages to and receive messages from other users, even when offline through the use of the storage-based, public key encrypted inbox.
 To save bandwidth and storage space, the files are not sent along with the message but are uploaded to the users' individual file storage and a link is attached to the message.
 Users are then able to send attachments using the file links from the Files Plugin.
\item
\textit{Wall}: This provides a 1-to-N communication.
    Each user views a personal wall pages that they can post entries to.
    Other users can also post to another user's personal page and also comment on those posts if they have been accepted as friends.
    The wall is one of the most complex plugin in terms of access control as various authors interact on a shared space.
\item
\textit{Photos}: Allows users to share photos and photo albums with other users and allows comments on the photos.
    Owners of photos/albums can choose whom they want to share with.
 \item
 \textit{Groups/Forum}: This great plugin models a computer supported cooperative work environment.
 When a group is created, the group of users are provided with a shared storage space to store files using the Files Plugin as well as a forum for posting discussions as well. 
 Due to the potentially huge storage space, complete working and interaction processes can be mapped in this. 
 Group admins can invite other users to the group. Through the concept of ``groups'' in the P2P framework, i.e. set of users, real-life working hierarchies can be mapped to corresponding working spaces (Group Plugin). 
\item
 \textit{Calendar}: The plugin allows users to store appointments and events chronologically.
 It utilizes a distributed data list to store the agenda of each user.
 Using the plugin the user can create, delete and modify events. 
 A collaborative sharing is envisioned. 
\item
 \textit{Voting}: Gives users the ability to conduct surveys among a predefined group of users or the entire set of users of the network.
 The vote initiator defines a question and a set of response choices, then invites other users to take part in the survey.
 Users can access other existing public votings, remove votes from their own vote list, invite others to vote and see the results from the votings after submitting their vote.
 \item
 \textit{Multichat}: Users can chat live with each other or in groups via this plugin through direct communication channels.
 The plugin lists the entire communication between a user and the conversation partner(s).
 It supports conversations between multiple users.
\item
 \textit{Audio/Video chat}: It provides an 1-to-1 audio and video communication through the usage of the WebRTC (https://webrtc.org) open standard.
 This is the only plugin that does not use the features of the P2P framework for communication but establishes individual browser-to-browser connections.
\item
 \textit{Monitoring}: This plugin performs statistical, aggregated monitoring of the entire system, on the application, the plugins, the framework and the overlay.
 Currently roughly 600 metrics are obtained and available to the users. 
 The monitoring plugin is based on the monitoring solution \textit{SkyEye.KOM}~\cite{GKX08,GAn17}.
\item
 \textit{Debug/Error Console}: This plugin is dedicated to displaying the information that is generated by the framework while the user is logged in, including information such as errors, warnings and debug messages.
\item
 \textit{Test}: This plugin is used to test and debug the performance and reliability of the entire application in a controlled environment. 
 For simple control of the testing, a centralized master-slave approach is used.
 The master sends the test plan to the slaves which execute it. 
 The test plan is a structured text document that contains a list of plugin commands in the given order and timing. 
 The results are obtained through the monitoring plugin and used to identify whether the preconditions for testing are met and other parameters are set correctly.
\end{itemize}

The plugins are divided into two categories, that is \textit{mandatory} and \textit{optional} plugins.
The plugins discussed above are all loaded into the system as mandatory plugins.
Further Apps loaded to the system can assume that these are present. 
Currently, App from the AppStore are considered as optional. 
Their dependencies must be resolved when installed, similarly as in the GNU packet environment.

\begin{figure*}[!tbp]
  \centering
  \begin{minipage}[b]{0.36\textwidth}
        \centering
        \includegraphics[scale=0.32]{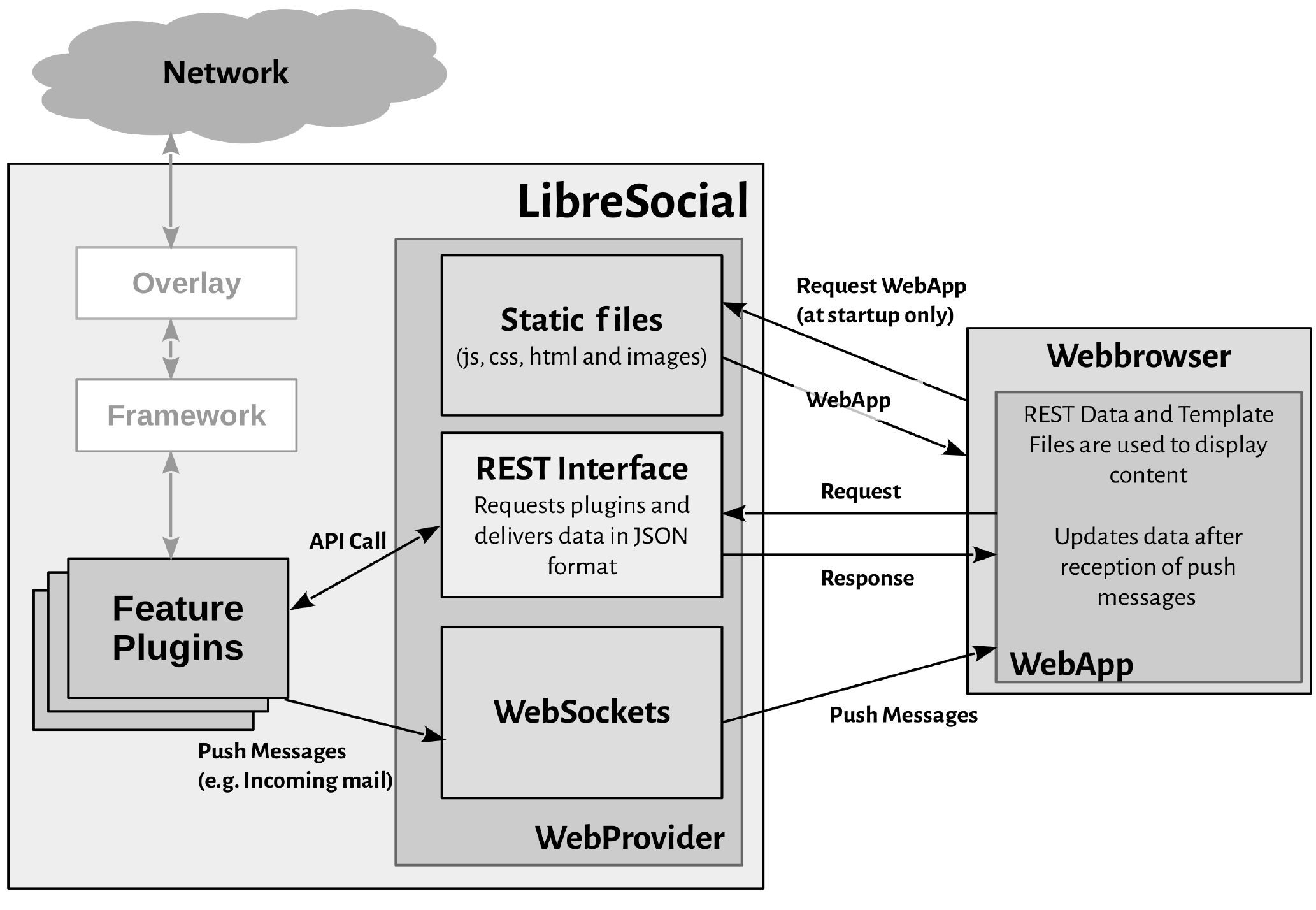}
        \caption{WebProvider}
        \label{fig:WebProvider}
  \end{minipage}
  \hfill
  \begin{minipage}[b]{0.63\textwidth}
        \centering    
        \includegraphics[height=3.2cm]{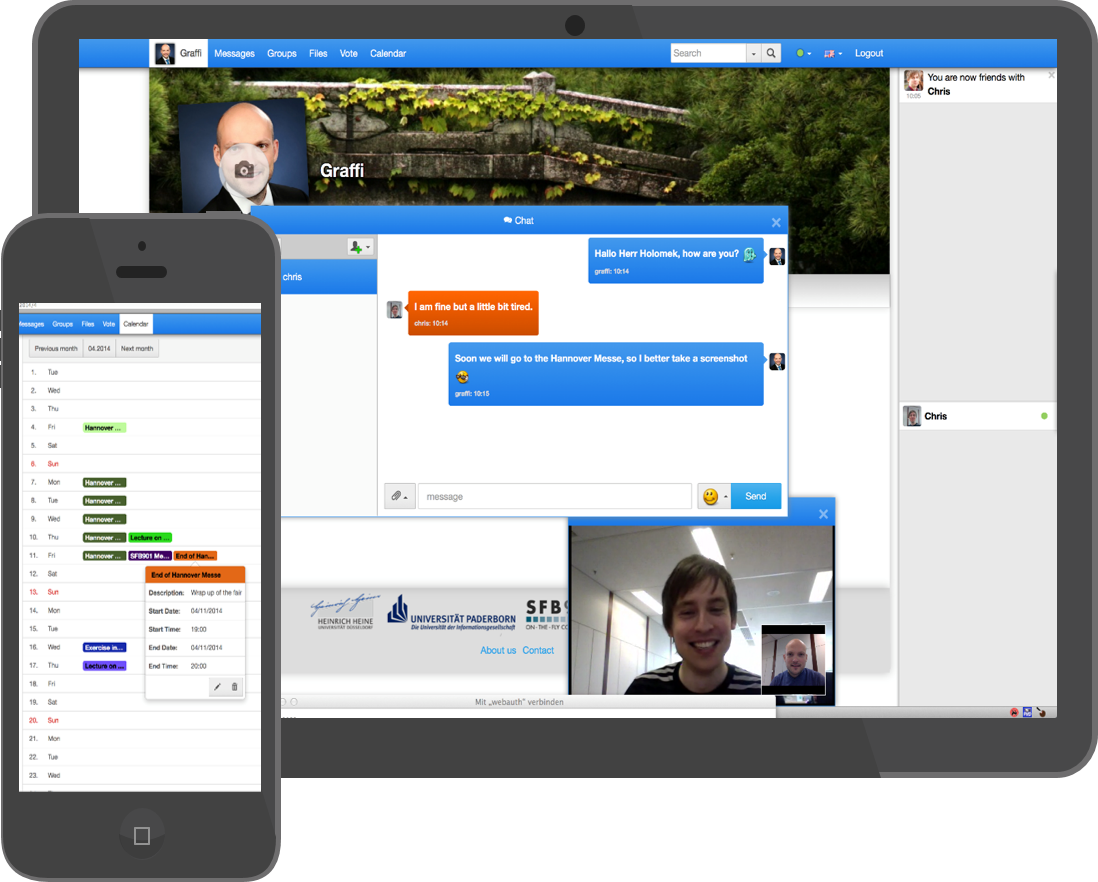}
        \includegraphics[height=3.2cm]{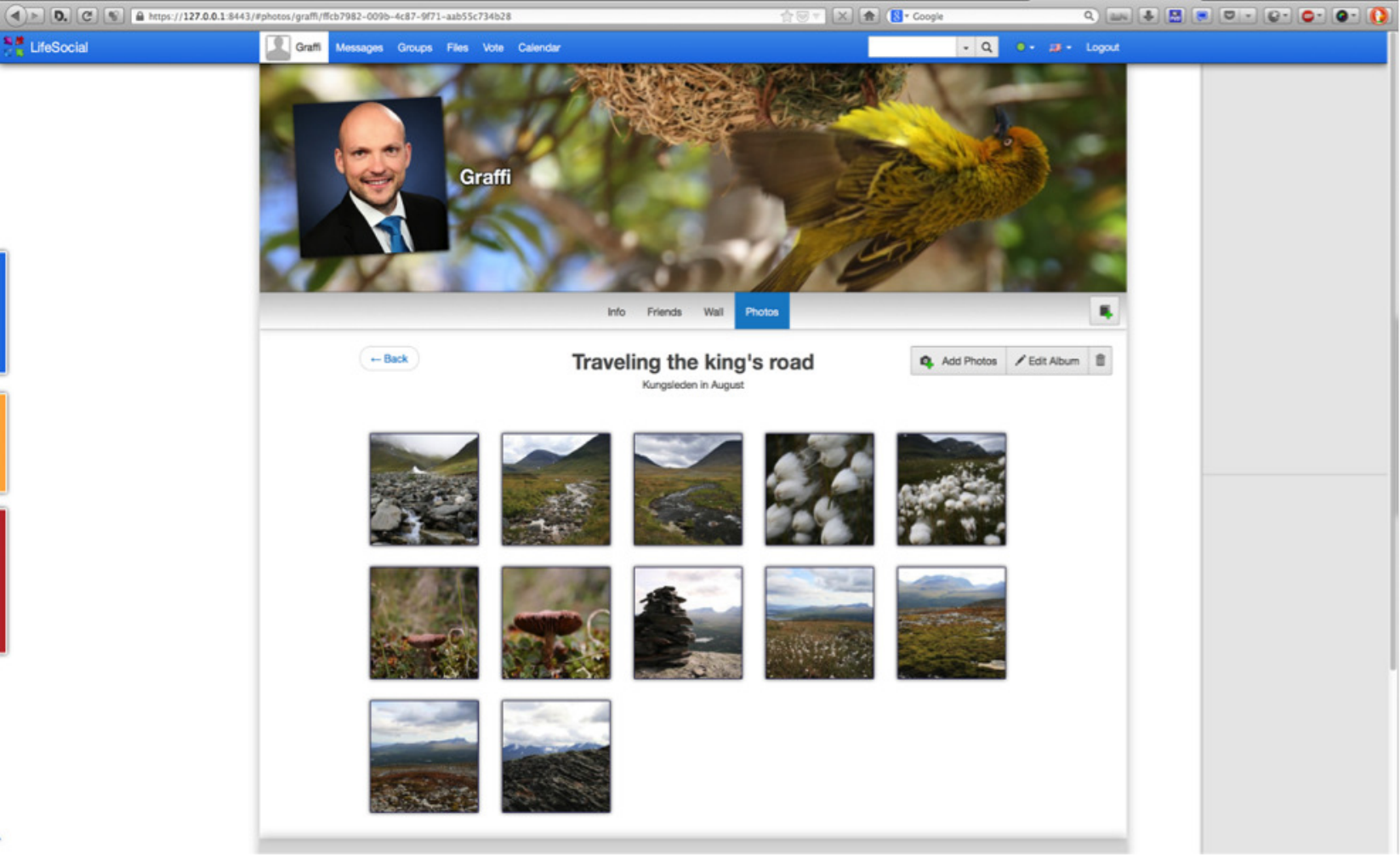}

        \includegraphics[height=3.2cm]{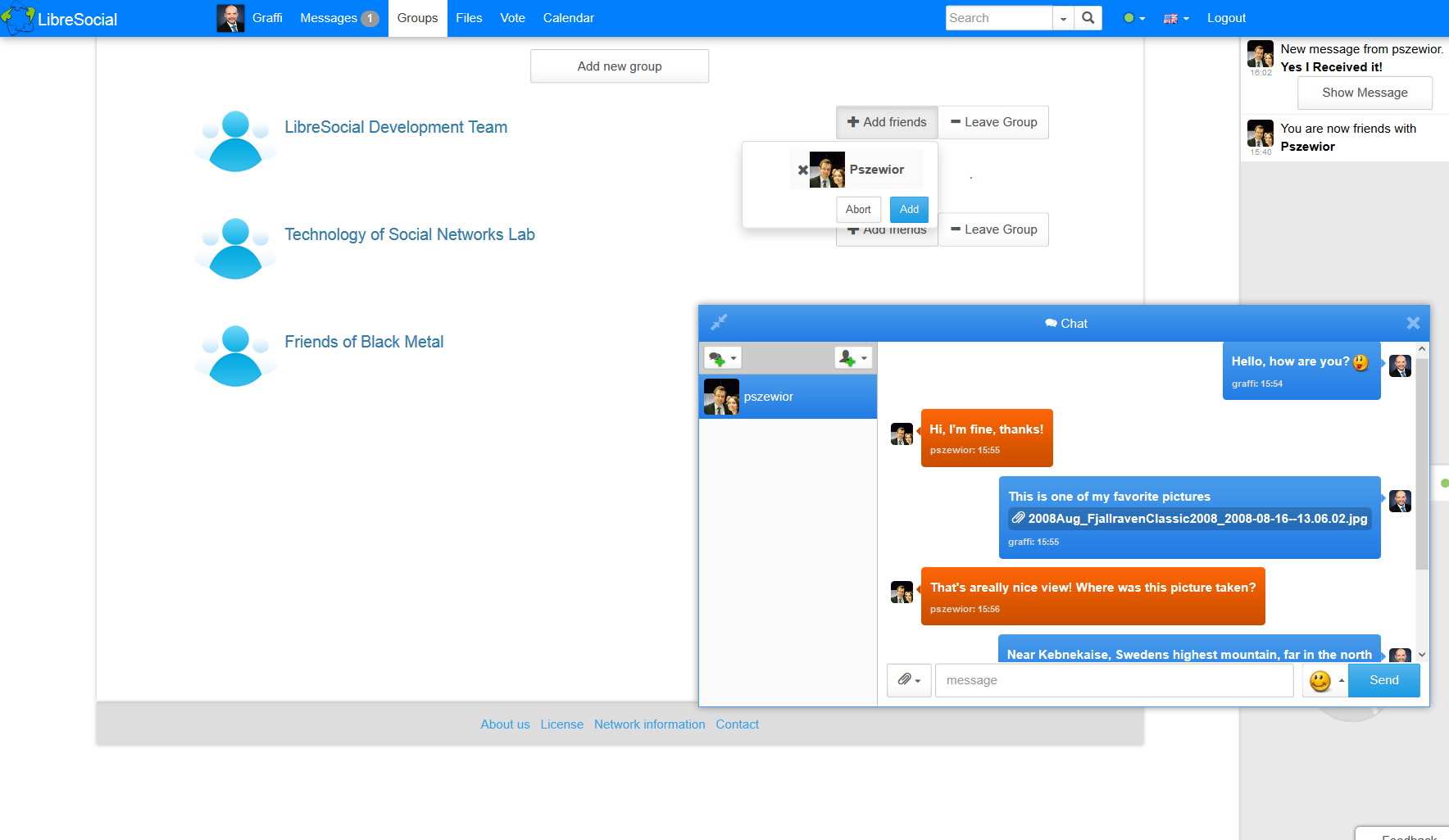}
        \includegraphics[height=3.2cm]{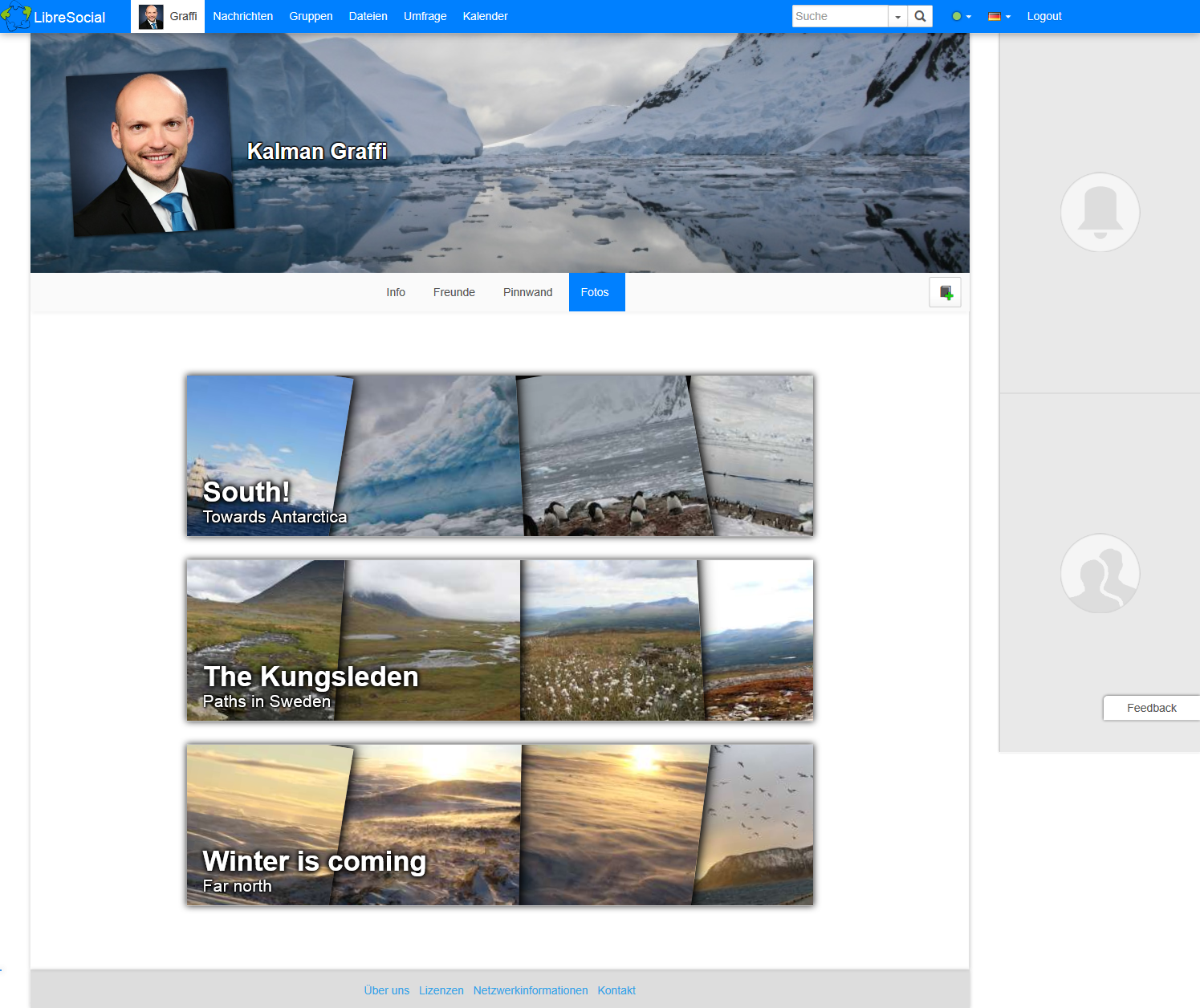}
        \caption{Screenshots of LibreSocial}
        \label{fig:LibreSocial_Screenshot}
  \end{minipage}
  \vspace{-0.4cm}
\end{figure*}

\section{The Graphical User Interface}
\label{sec:GUI}
The top most layer of the architecture (Fig.~\ref{fig:p2p_framework_overview}) is the graphical user interface (GUI), which is the point of contact with the users.
LibreSocial's GUI has undergone many changes since the initial implementation.
Starting from a pure command line interface, to an Eclipse-based applet framework, to its current design which uses standard web technologies such as HTML5, AJAX, JQuery, Bootstrap and Knockback.js.
The current GUI is shown as a screenshot in Fig.~\ref{fig:LibreSocial_Screenshot}.
This combination of web technologies also allows the provision of multi-language support (currently only English and German) as well as support for mobile devices (at least the GUI is accessible).
The GUI's  backend is composed of three essential parts, the \textit{plugin template}, the \textit{plugin logic} and the \textit{WebProvider}.
These are discussed in detail.

\textit{The plugin template} are typical HTML files which use Knockout.js (knockoutjs.com), a standalone JavaScript implementation based on the Model-View-View-Model (MVVM) paradigm.
Knockout.js is small and very light weight and presents several advantages in particular, declarative binding of elements to models, dependency tracking to reflect values when the dependency changes, as well as automatic user interface refresh when the data model's state changes.

\textit{The plugin logic} deals with the entire process of transferring user events from the frontend to the REST Handler via the WebProvider and back to the frontend again.
The plugin logic renders data it receives to the desired template.
The proper functioning of this mechanism requires the creation of models, collections and views, which is contrary to the MVVM paradigm that provides only models and views.
To support the creation of Models and collections, Backbone.js (backbonejs.org) is used, and to support views, Knockback.js (kmalakoff.github.io/knockback) is used.
The use of Backbone.js gives structure to web applications as it provides the models with \texttt{key-value} binding and custom events.
The data in represented as models with Backbone.js, which makes it flexible to create, validate and destroy the models based on the system designer's needs.
In case a change is made to an attribute of the model through the user interface, a change event is fired back which can be used by the view to remodify itself based on the model modification.

The \textit{WebProvider} is a plugin that functions as the interface between plugins and the GUI and uses the embeddable web server jetty (www.eclipse.org/jetty) to communicate with the user's browser, which acts as the frontend.
Despite the fact that this jetty module uses a single network port, it can handle three separate tasks distinguishable by the request URLs.
Fig.\ref{fig:WebProvider} is the structure of the WebProvider.
The three tasks, provision of the static files, the REST interface and the WebSockets, constitute the backend.

The browser is the frontend, and does not simply present the information emanating from a server, but rather runs as a JavaScript web application, while managing data exchange with the actual P2P application.
The views are constructed by bringing together data from the feature plugins and the corresponding templates.
With a few exceptions, such as the login screen, the browser page is never reloaded entirely but only relevant parts are updated or exchanged between other modules.
This presents some advantages for social network applications such as LibreSocial, where communication is desired to be in real time.
This means that several interaction objects can be opened simultaneously and updated individually.

Although it is technically not possible to install the LibreSocial application on mobile devices because of bandwidth and availability limitations, it is nevertheless possible to use the GUI on mobile devices.
Therefore, to support future use of LibreSocial while on the go, a user interface that works for mobile devices is implemented.
\textit{Bootstrap} (getbootstrap.com) is used to provides GUI elements as building blocks that are automatically organized so as to optimize the screen size and it also ensures the interaction controls, such as buttons or input fields, are large enough for touch screen operation.
To ensure that the connection between the backend and the frontend of the application is secure, even if both parts are domiciled in the same computer, the connection is transport encrypted using normal Transport Layer Security/Secure Sockets Layer (TLS/SSL) based HTTPS as opposed to using plain HTTP.

\section{LifeSocial.KOM vs.  LibreSocial}
\label{sec:Comparison}
In this section, we give an analytical comparison of the features of LifeSocial.KOM and LibreSocial, and show where there are significant changes in the components or the functionalities.
In Table~\ref{tab:AppDifferences}, the differences are tabulated.
We give a summary of the differences below.
\begin{itemize}
\item
 \textit{Identity management}: Due to the change from using a 1024-bit RSA algorithm to a 160-bit ECC algorithm as the PKI, the node ID space is changed from a 1024-bit ID space to a 160-bit ID space.
 This allows the generated public key to be directly used in the network for encrypted communication.
\item
 \textit{Message routing strategy}: In addition to the traditional forwarding strategy to the node with \texttt{nodeID} that is numerically closest to the required \texttt{nodeID}, LibreSocial also implements a parallel/iterative routing strategy.
 This mitigates against certain attacks such as Sybil, Eclipse and routing attacks, while also limiting the lookup time and the amount of traffic generated.
\item
 \textit{Capacity awareness}: By supporting heterogeneous nodes in LibreSocial, it is possible to further introduce strategies that allow the system to gather more informative data such as available persistent storage space, memory, bandwidth, and type of devices at login and even at runtime so as to dynamically adjust the routing table to support strong and weak nodes.
\item
 \textit{Group access control}: A group access control mechanism has been introduced to support sets and nested sets of users allowing to ease the management of various friend groups and to support the mapping of organizational hierarchies to e.g. the CSCW groups and forums. 
\item
 \textit{Distributed data structures}: In addition to the use of the distributed set/linked list available in LifeSocial.KOM, LibreSocial also includes Prefix Hash Trees which allows range queries. 
 This is used when searching for other users based on a given range of values such as name, location and so on. 
 This feature could also be used in other future plugins. 
\item
 \textit{Live conferencing}: LifeSocial did not include support for audio/video live conferencing.
 This has been included in LibreSocial using WebRTC supporting all modern browsers.
\item
 \textit{Additional plugins}: LibreSocial includes plugins that extend functionality of the OSN  not present in LifeSocial.KOM namely Audio/Video Chat, Wall, Forums, Voting and an Error Console.
\item
 \textit{Graphical User interface}: As opposed to the Eclipse window-based interface design in LifeSocial.KOM, LibreSocial implements a web-based user interface that is easier and has more appeal to the general user.
\end{itemize}

\begin{figure*}[!tbp]
  \centering
  \begin{minipage}[t]{0.63\textwidth}
        \small \centering
 \captionof{table}{Comparison between LifeSocial.KOM and LibreSocial}
 \vspace{-0.3cm}
 \label{tab:AppDifferences}
 \begin{tabular}{|p{0.9cm}|p{0.8cm}|p{1.7cm}|p{2.5cm}|p{3cm}|}
  \hline \textbf{Architectural Level} &
  \textbf{Functionality} &
  \textbf{Feature} &
  \textbf{LifeSocial.KOM} &
  \textbf{LibreSocial} \\
  \hline \rule{0pt}{2ex}\textbf{Overlay} &
  \textit{Identity } &
  Identity Space &
  1024-bit nodeID &
  160-bit nodeID \\
  \cline{2-5} \rule{0pt}{2ex} &
  \textit{Messaging} &
  Routing performance &
  O(logN) &
  O(logN) \\
  \cline{3-5} \rule{0pt}{2ex} &
  &
  Routing strategy &
  Query forwarding &
  Query forwarding.
  Supports parallel and iterative routing \\
  \cline{2-5} \rule{0pt}{2ex} &
  \textit{Capacity awareness} &
  Mobile vs static nodes &
  Not supported &
  Statically supported \\
  \cline{2-5} \rule{0pt}{2ex} &
  \textit{Security} &
  Asym. encryption &
  RSA with 1024-bit keys &
  ECC with 160-bit keys \\
  \cline{3-5} \rule{0pt}{2ex} &
  &
  Symmetric encryption &
  AES with 128-bit key &
  AES with 128-bit key \\
  \hline \rule{0pt}{2ex}\textbf{Framework} &
  \textit{Storage} &
  File management &
  PAST &
  Heavily extended PAST \\
  \cline{3-5} \rule{0pt}{2ex} &
  and &
  File access control &
  Supported &
  Supported \\
  \cline{3-5} \rule{0pt}{2ex} &
  replication &
  Group access control &
  Not supported &
  Supported \\
  \cline{3-5} \rule{0pt}{2ex} &
  &
  Distributed data structures &
  Distributed set, distributed linked list &
  Distributed set, distributed linked list, prefix hash trees \\
  \cline{2-5} \rule{0pt}{2ex} &
  \textit{Communi-} &
  Publish-subscribe &
  Scribe &
  Scribe \\
  \cline{3-5} \rule{0pt}{2ex} &
  \textit{cation} &
  Multicast streaming &
  SplitStream &
  SplitStream \\
  \cline{3-5} \rule{0pt}{2ex} &
  &
  Live conferencing &
  Not supported &
  Implemented using WebRTC \\
  \cline{2-5}
  \rule{0pt}{2ex} & Optional & Monitoring & Tree-based monitoring (SkyEye) &
  Tree-based monitoring (SkyEye) \\
  \hline \rule{0pt}{2ex}\textbf{Plugin and application} &
  Plugins &
  Application plugins &
  Login, Profile, Notification, Files, Search, Friends, Groups, Calendar, Messaging, Multichat, Photos, Testing, Monitoring.
  &
  Login, Profile, Notification, Files, Search, Friends, Groups, Calendar, Messaging, Multichat, Photos, Testing, Monitoring, Error Console, Audio/Video chat, Wall, Forums, Voting \\
  \cline{3-5} \rule{0pt}{2ex} &
  &
  User interface &
  Window-based &
  Web-based\\
  \hline
 \end{tabular}
\begin{minipage}[t]{0.50\textwidth}
\vspace{0.3cm}
 \small \centering
 \caption{Evaluation of ECC \\ - the asymmetric algorithm as PKI}
 \vspace{-0.3cm}
 \label{tab:Asymmetric}
 \begin{tabular}{|P{0.30cm}P{0.4cm}P{0.4cm}P{0.55cm}|P{0.2cm}P{0.3cm}P{0.1cm}P{0.25cm}|}
  \hline \multicolumn{4}{|c|}{\textbf{Data Size (bytes)}} &
  \multicolumn{4}{c|}{\textbf{Time taken (ms)}} \\ 
  \hline \rule{0pt}{2ex}\textit{Original} & \textit{Encrypted} &
  \textit{Overhead} &
  \textit{Signature} &
  \textit{Encrypt} &
  \textit{Decrypt} &
  \textit{Sign} &
  \textit{Verify} \\
  \hline \rule{0pt}{2ex}1455 &
  1497 &
  42 &
  48 &
  7.75 &
  2.93 &
  2.79 &
  3.81 \\
  \rule{0pt}{2ex}1461 &
  1513 &
  52 &
  47 &
  6.35 &
  3.24 &
  1.90 &
  3.39 \\
  \rule{0pt}{2ex}1493 &
  1545 &
  52 &
  47 &
  3.33 &
  3.73 &
  2.81 &
  8.07 \\
  \rule{0pt}{2ex}1807 &
  1849 &
  42 &
  47 &
  11.52 &
  4.93 &
  4.58 &
  7.68 \\
  \rule{0pt}{2ex}2013 &
  2057 &
  44 &
  47 &
  9.54 &
  12.94 &
  2.72 &
  4.90 \\
  \rule{0pt}{2ex}2133 &
  2185 &
  52 &
  48 &
  3.22 &
  2.82 &
  1.61 &
  3.19 \\
  \hline
 \end{tabular}
 \end{minipage}
 \hfill
 \begin{minipage}[t]{0.45\textwidth} 
 \vspace{0.3cm}
 \tiny \centering
 \caption{Evaluation of AES \\ - the symmetric encryption algorithm}
 \vspace{-0.3cm}
 \label{tab:Symmetric}
 \begin{tabular}{|P{0.3cm}P{0.4cm}P{0.6cm}|P{0.3cm}P{0.5cm}|}
  \hline \multicolumn{3}{|c|}{\textbf{Data size (bytes)}} &
  \multicolumn{2}{c|}{\textbf{Time (ms)}} \\
  \hline \rule{0pt}{2ex}\textit{Original} & \textit{Encrypted} &
  \textit{Overhead} &
  \textit{Encrypt} &
  \textit{Decrypt} \\
  \hline \rule{0pt}{2ex}609 &
  624 &
  15 &
  0.13 &
  0.44 \\
  \rule{0pt}{2ex}784 &
  800 &
  16 &
  0.12 &
  0.12 \\
  \rule{0pt}{2ex}805 &
  816 &
  11 &
  0.09 &
  0.09 \\
  \rule{0pt}{2ex}827 &
  832 &
  5 &
  0.20 &
  0.29 \\
  \rule{0pt}{2ex}845 &
  848 &
  3 &
  0.11 &
  0.14 \\
  \rule{0pt}{2ex}880 &
  896 &
  16 &
  0.11 &
  0.21\\
  \hline
 \end{tabular}
  \end{minipage}
  \end{minipage}
  \hfill
  \begin{minipage}[t]{0.36\textwidth}
\centering \small
 \captionof{table}{Baseline test workload}
 \vspace{-0.3cm}
 \label{tab:baseline_workload}
 \begin{tabular}{|l|l|c|c|}
  \hline Plugin &
  Action &
  \begin{tabular}[c]{@{}c@{}}Repetitions\\
   (Count)\end{tabular} &
  \begin{tabular}[c]{@{}c@{}}Duration \\
   (mins)\end{tabular} \\
  \hline \rule{0pt}{2ex}Messaging &
  Send message &
  8 &
  2 \\
  &
  View inbox messages &
  4 &
  2 \\
  &
  View outbox messages &
  4 &
  2 \\
  \hline \rule{0pt}{2ex}Livechat &
  Send multichat invitation &
  2 &
  1 \\
  &
  Send multichat message &
  8 &
  2 \\
  &
  Leave multichat message &
  2 &
  1 \\
  \hline \rule{0pt}{2ex}Group &
  Create group &
  4 &
  2 \\
  &
  Invite friend to group &
  4 &
  2 \\
  &
  View group &
  8 &
  2 \\
  &
  Leave group &
  2 &
  2 \\
  \hline \rule{0pt}{2ex}Filestorage &
  Create folder &
  4 &
  2 \\
  &
  Upload file in folder &
  32 &
  6 \\
  &
  View folder &
  4 &
  2 \\
  &
  Delete file from folder &
  8 &
  2 \\
  &
  Delete folder &
  2 &
  2 \\
  \hline \rule{0pt}{2ex}Forum &
  Create forum thread &
  4 &
  2 \\
  &
  Comment forum thread &
  4 &
  2 \\
  &
  View forum thread &
  4 &
  2 \\
  &
  View forum &
  4 &
  2 \\
  \hline \rule{0pt}{2ex}Photos &
  Create photo album &
  4 &
  2 \\
  &
  Upload photo &
  32 &
  6 \\
  &
  View own album &
  4 &
  2 \\
  &
  View friend's album &
  4 &
  2 \\
  &
  View friend's photo &
  32 &
  6 \\
  &
  Delete photo &
  16 &
  2 \\
  &
  Delete album &
  4 &
  2 \\
  \hline \rule{0pt}{2ex}Voting &
  Create vote &
  4 &
  2 \\
  &
  Add public vote &
  4 &
  2 \\
  &
  Voting invite user &
  4 &
  2 \\
  &
  Vote &
  16 &
  4 \\
  &
  Get my votings &
  4 &
  2 \\
  &
  Get voting results &
  16 &
  2 \\
  \hline \rule{0pt}{2ex}Wall &
  Send wall post &
  4 &
  2 \\
  &
  Comment wall post &
  4 &
  2 \\
  &
  View own wall &
  4 &
  2 \\
  &
  View friend's wall &
  4 &
  2 \\
  \hline
 \end{tabular}
  \end{minipage}
  \vspace{-0.4cm}
\end{figure*}

\begin{figure*}[htbp!]
 \centering 
 \subfloat[Message plugin: view inbox count]{\label{fig:PLUGIN_MESSAGING_VIEWINBOX}\includegraphics[width = 1.7in]{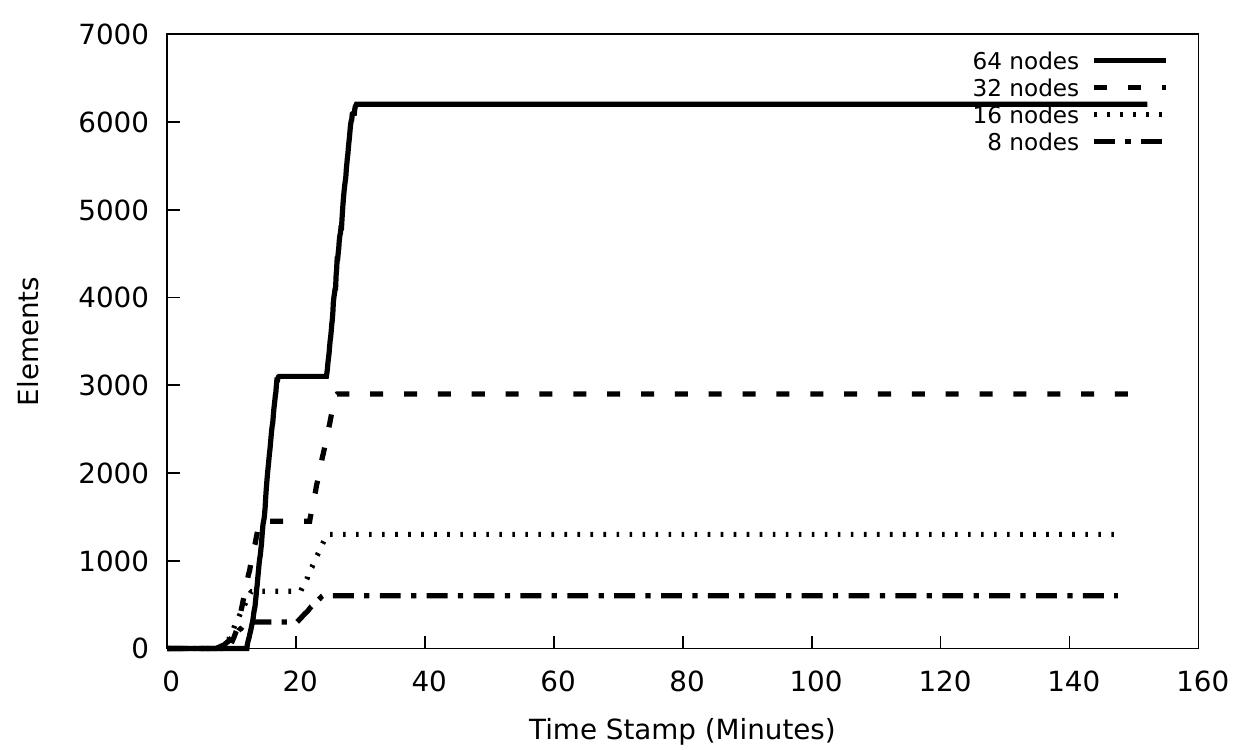}}
 \subfloat[Friends plugin: friendship requests]{\label{fig:PLUGIN_FRIENDS_FRIENDSHIPREQUEST}\includegraphics[width = 1.7in]{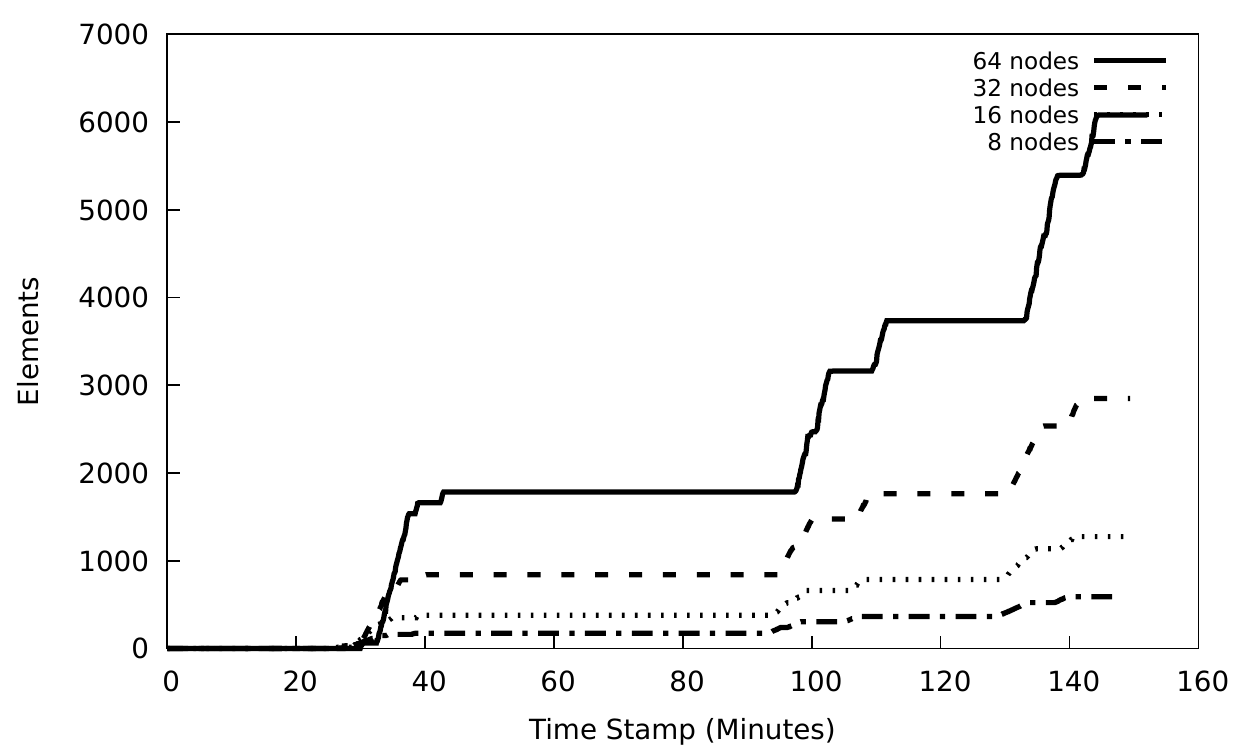}} 
 \subfloat[Group plugin: group view requests]{\label{fig:PLUGIN_GROUPS_VIEWGROUP}\includegraphics[width = 1.7in]{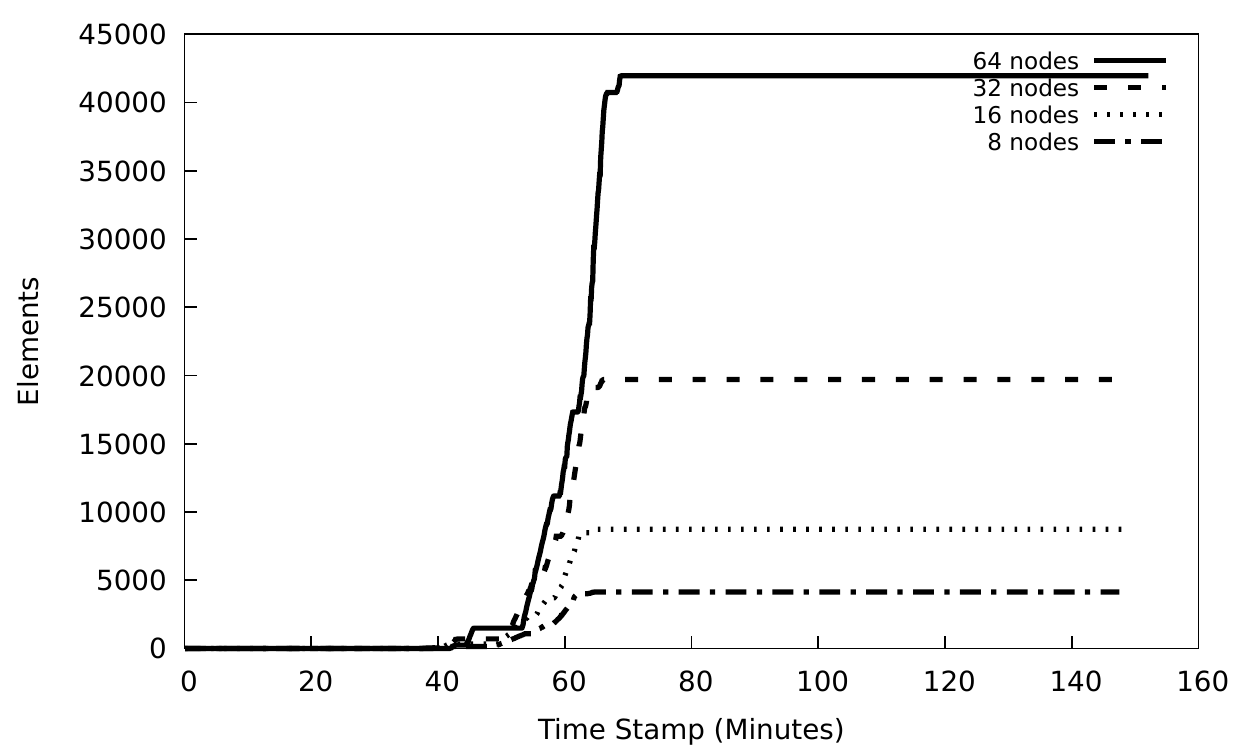}} 
 \subfloat[Photo plugin: albums created]{\label{fig:PLUGIN_PHOTOS_CREATEDALBUM}\includegraphics[width = 1.7in]{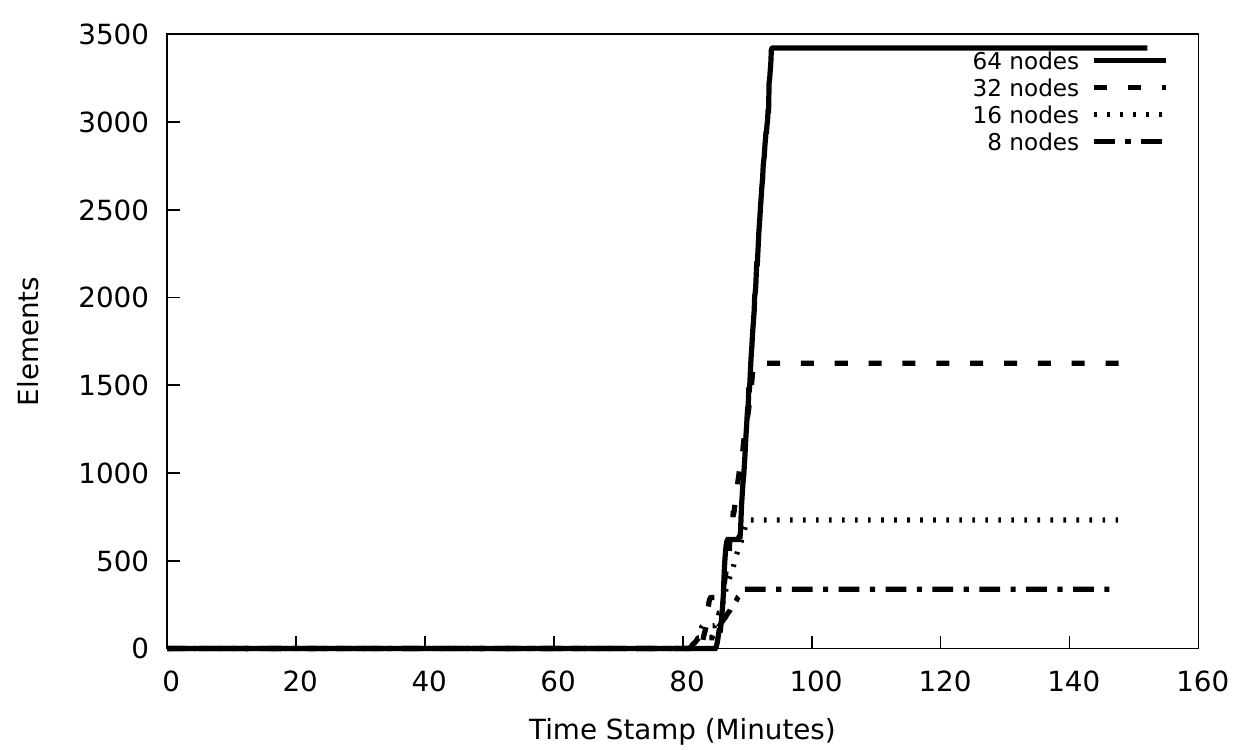}}
 \\
\vspace{-0.3cm}
 \subfloat[Total data receive rate ]{\label{fig:NETWORK_DATA_READ_RATE}\includegraphics[width = 1.7in]{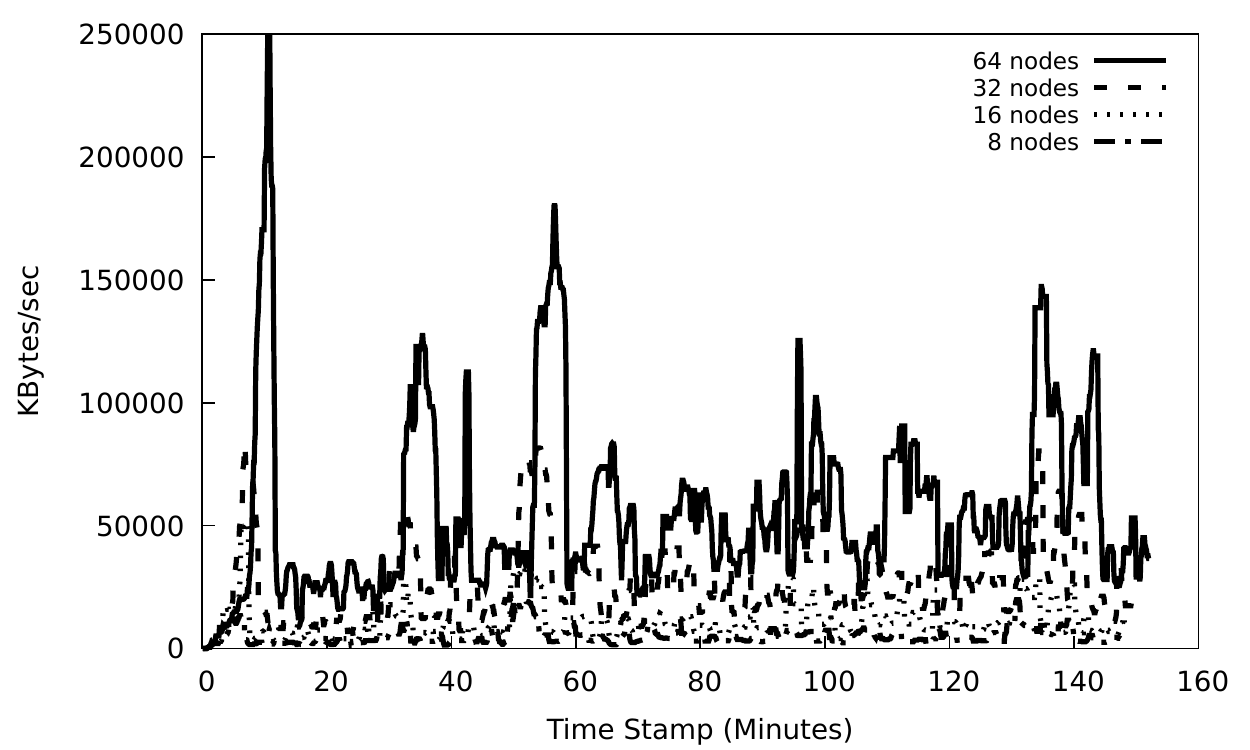}} 
 \subfloat[Max. data receive rate per node]{\label{fig:PASTRY_DATA_READ_RATE}\includegraphics[width = 1.7in]{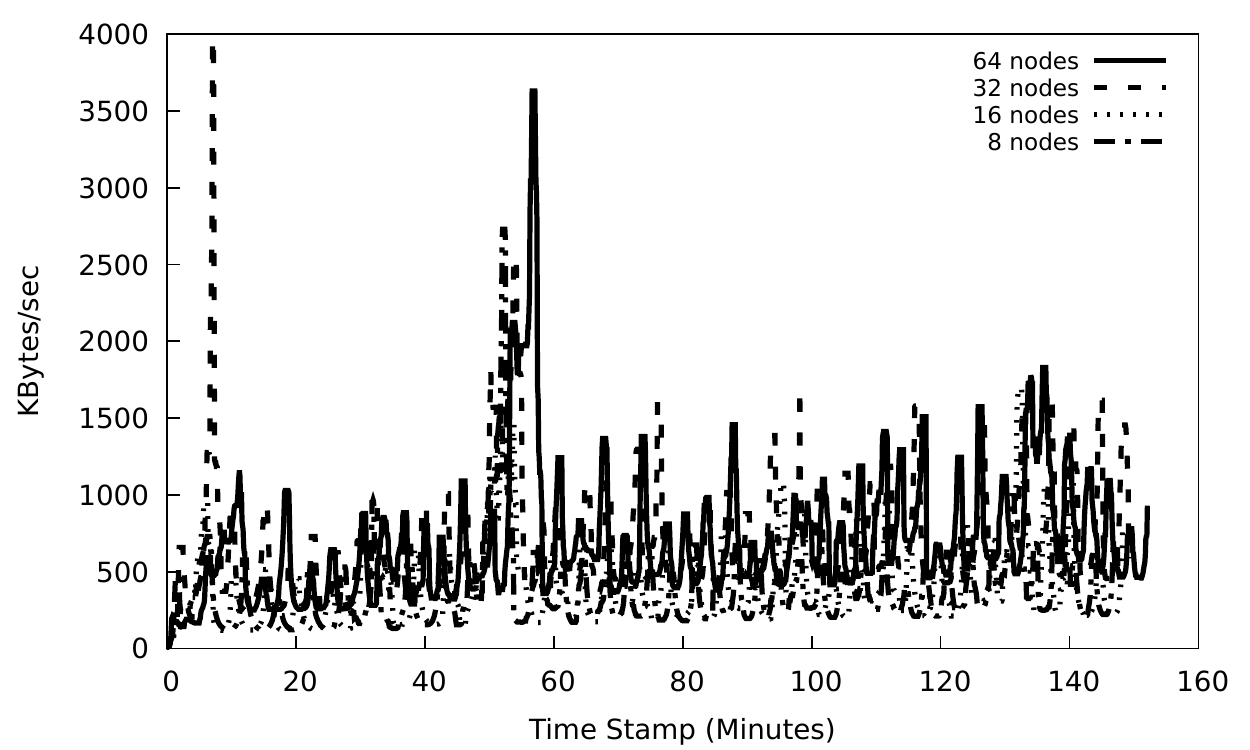}}
 \subfloat[Max. DDS retrieve rate per node]{\label{fig:STORAGE_RETRIEVED_DDS_DATA_RATE_MAX}\includegraphics[width = 1.7in]{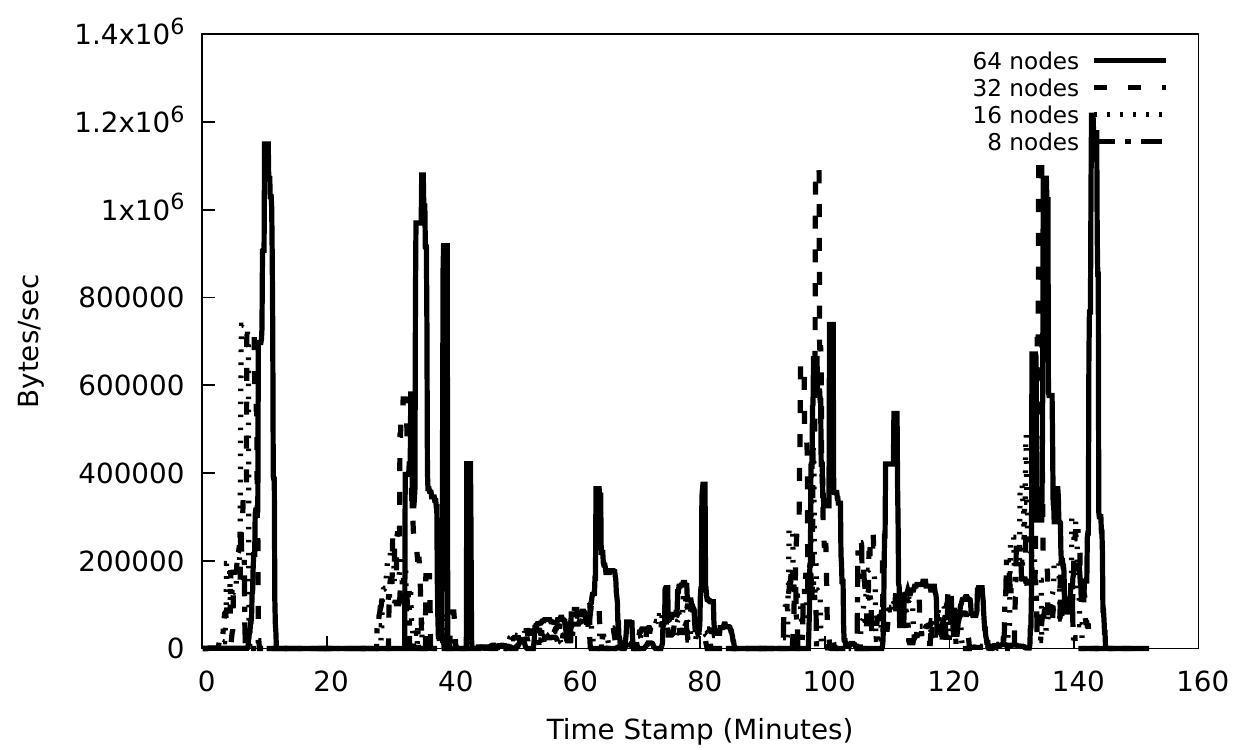}}
 \subfloat[Total message receive rate]{\label{fig:MESSAGES_RECEIVED_RATE}\includegraphics[width = 1.7in]{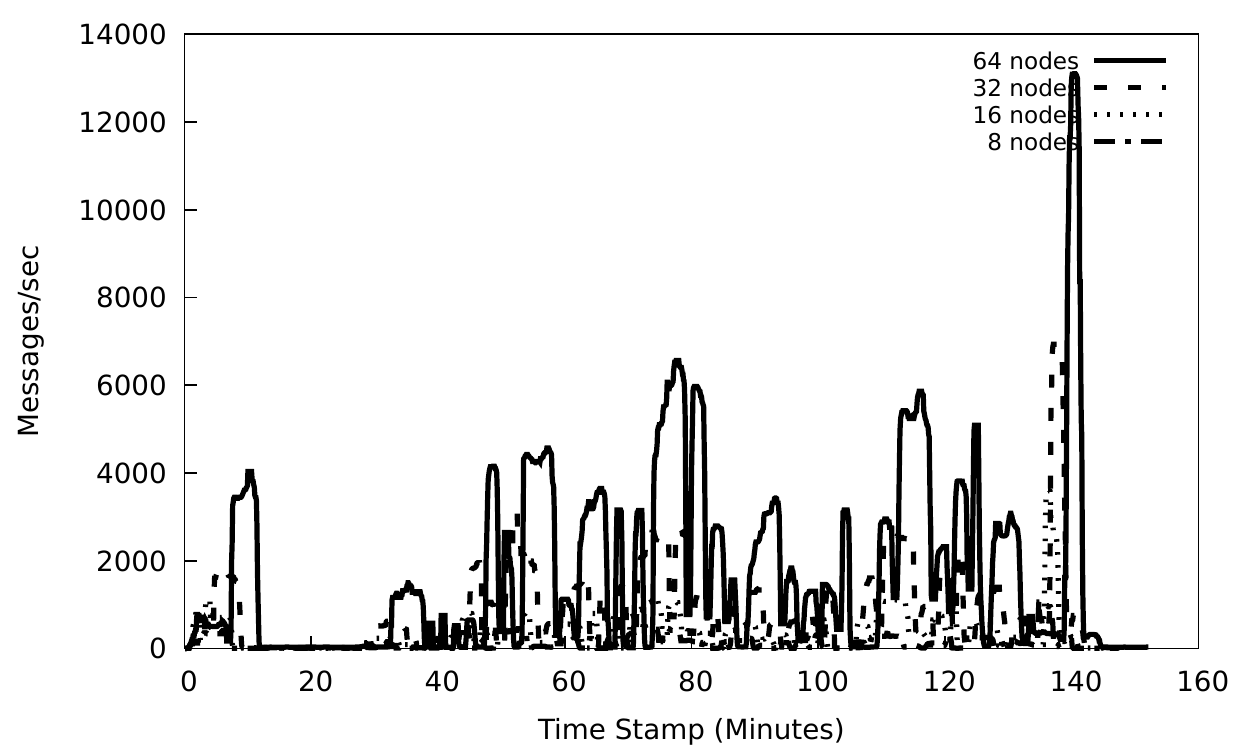}}
\\
\vspace{-0.3cm}
 \subfloat[Average message hop count]{\label{fig:MESSAGE_HOPCOUNT_MEAN}\includegraphics[width = 1.7in]{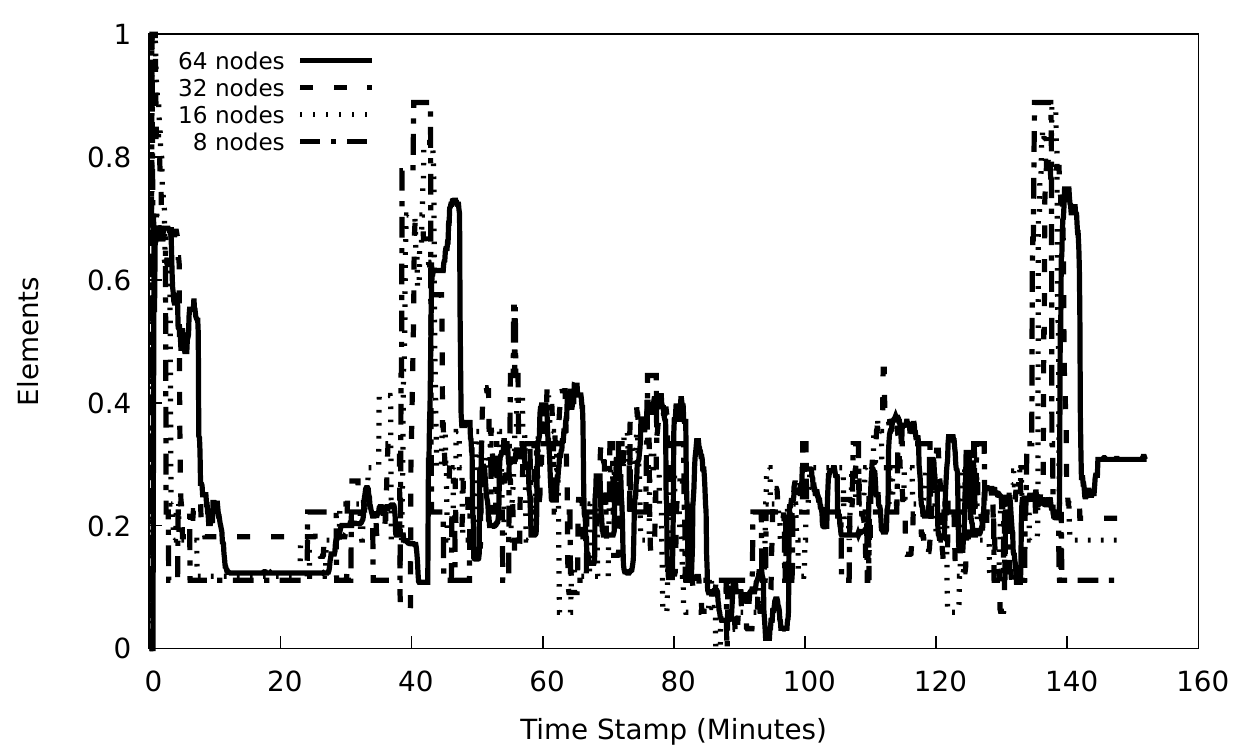}} 
 \subfloat[Average retrieval time]{\label{fig:RETRIEVETIME_MEAN}\includegraphics[width = 1.7in]{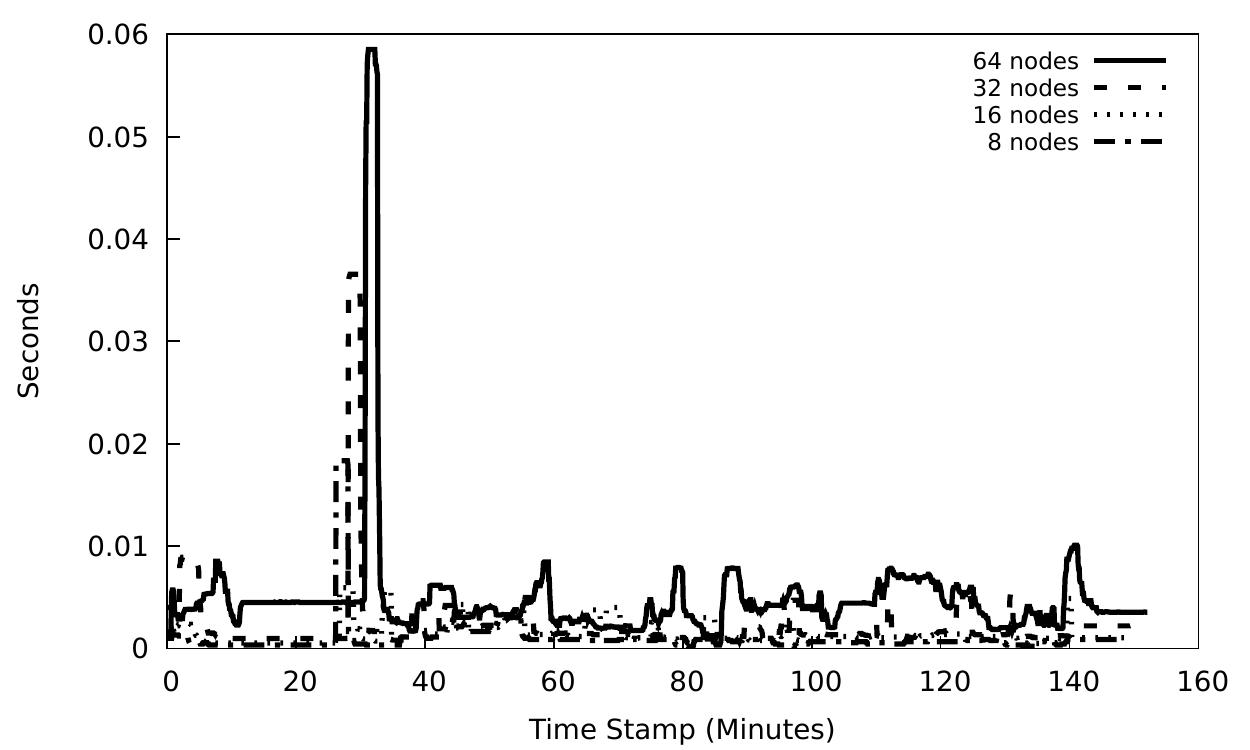}} 
 \subfloat[Average storage time]{\label{fig:STORETIME_MEAN}\includegraphics[width = 1.7in]{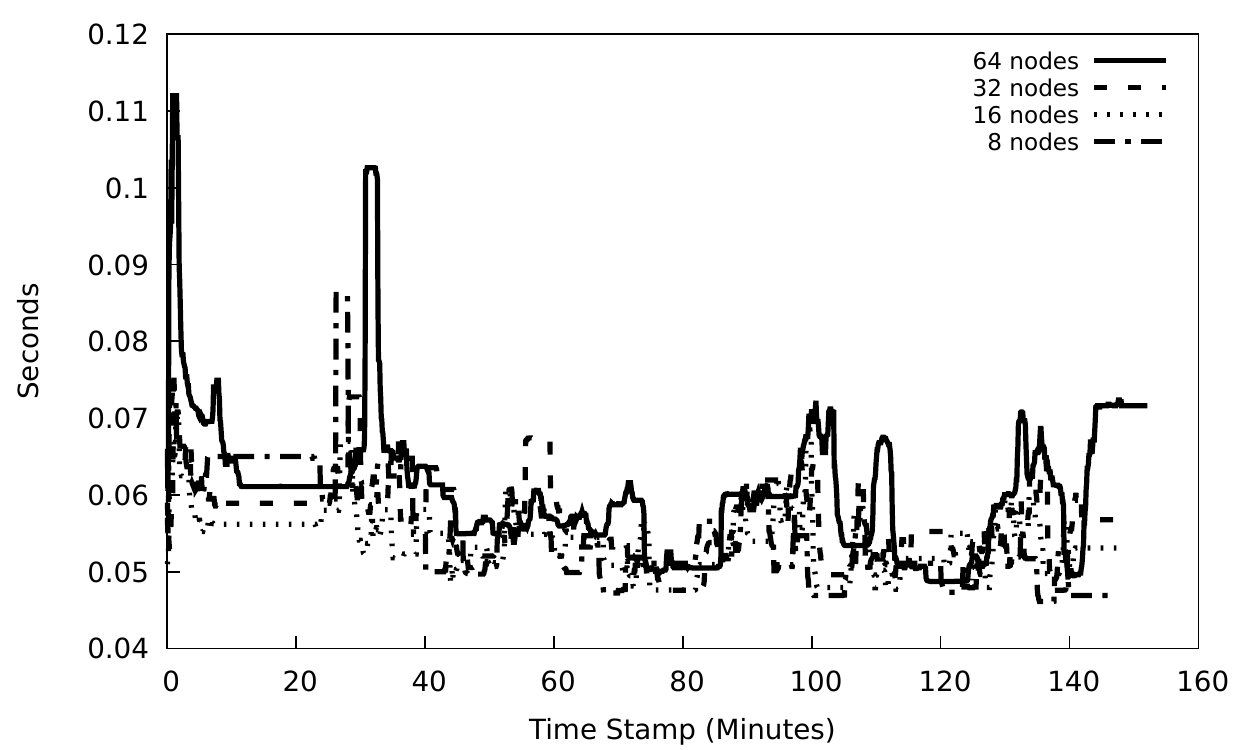}}
 \subfloat[Max. message receive rate per node]{\label{fig:MESSAGES_RECEIVED_RATE_MAX}\includegraphics[width = 1.7in]{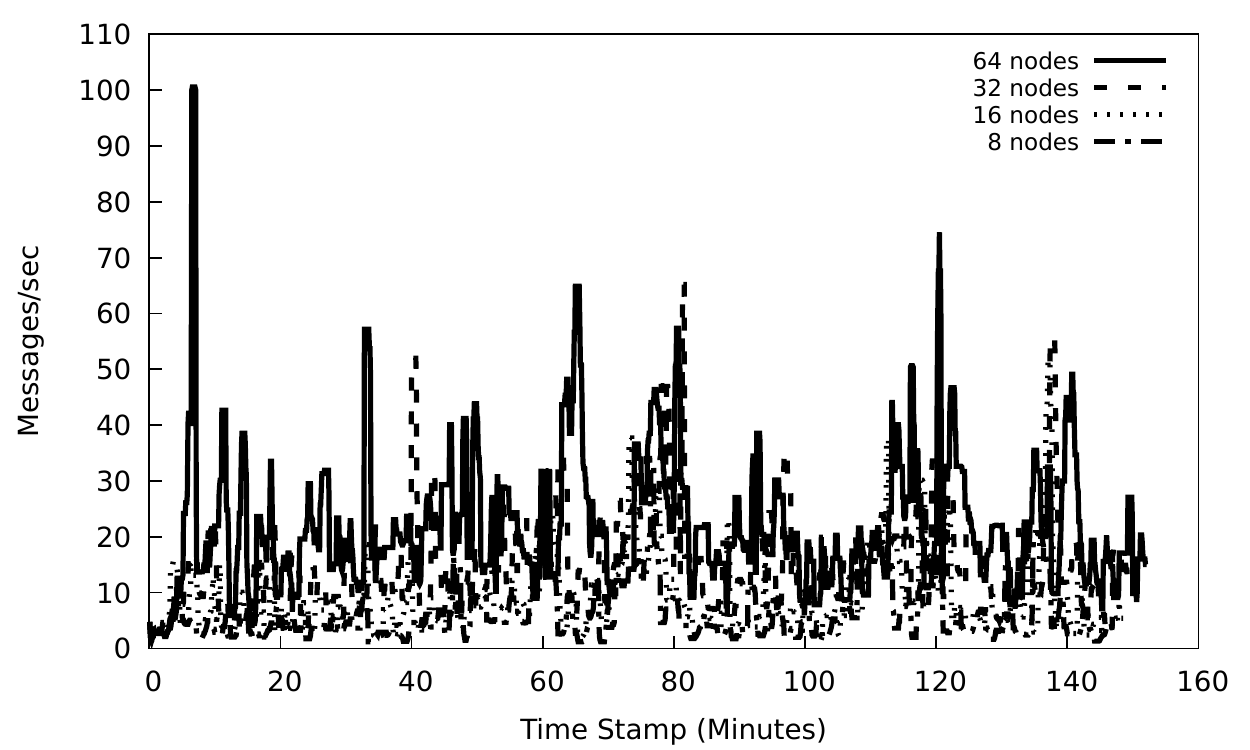}} 
 \\
\vspace{-0.3cm}
 \subfloat[Total data objects stored]{\label{fig:NUM_STORED_LOCAL_OBJECTS}\includegraphics[width = 1.7in]{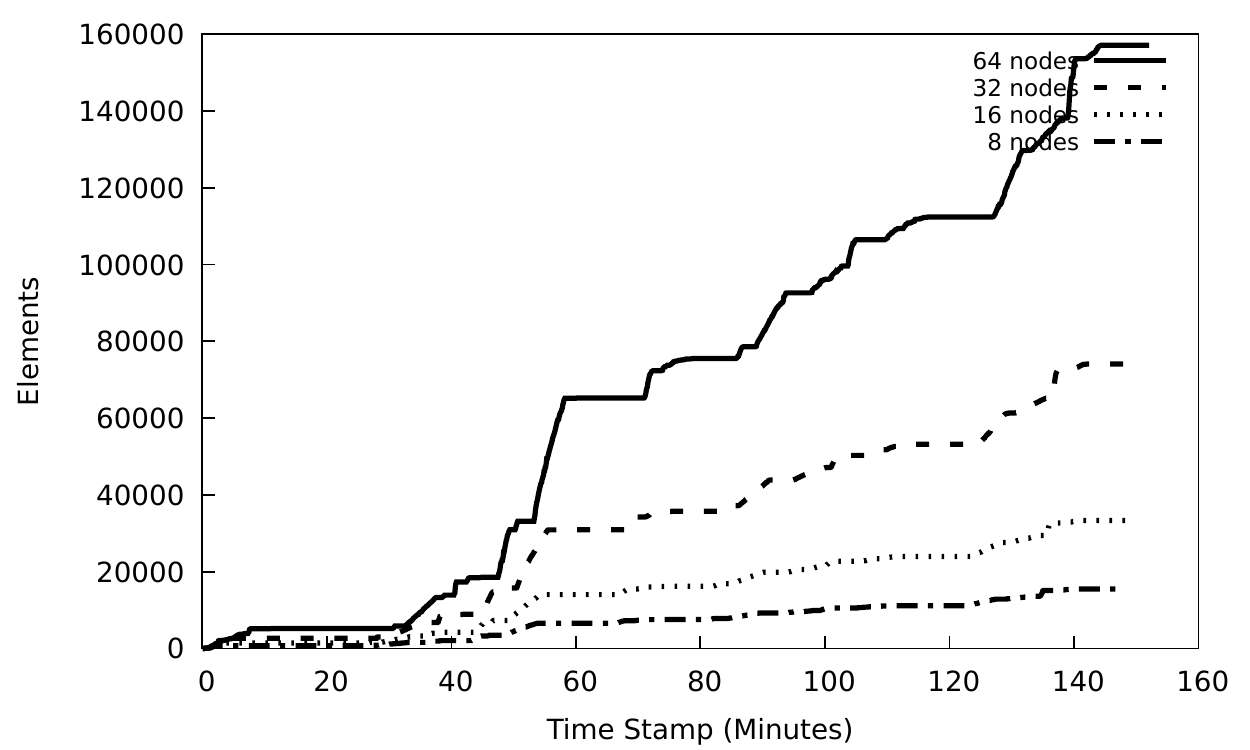}} 
 \subfloat[Max. data object count per node]{\label{fig:NUMINSTORAGE_ELEMENTS_MAX}\includegraphics[width = 1.7in]{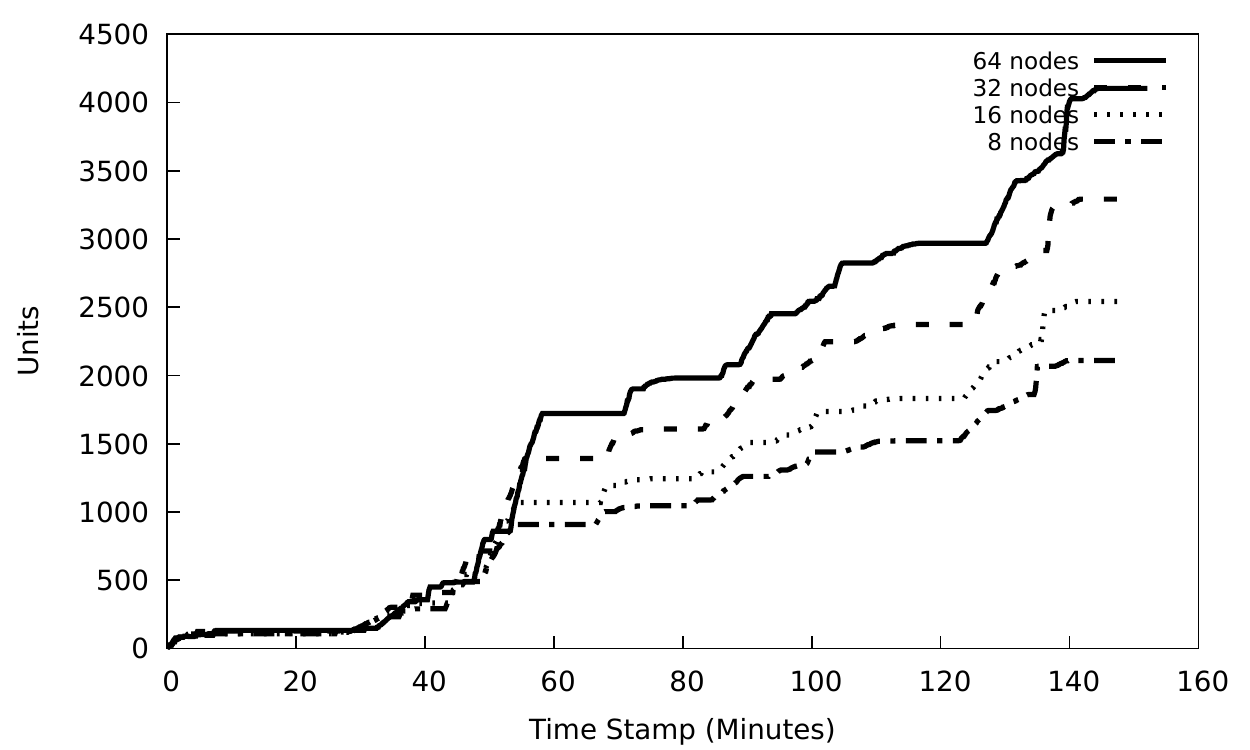}} 
 \subfloat[Total replica count]{\label{fig:NUM_STORED_REPLICATIONS}\includegraphics[width = 1.7in]{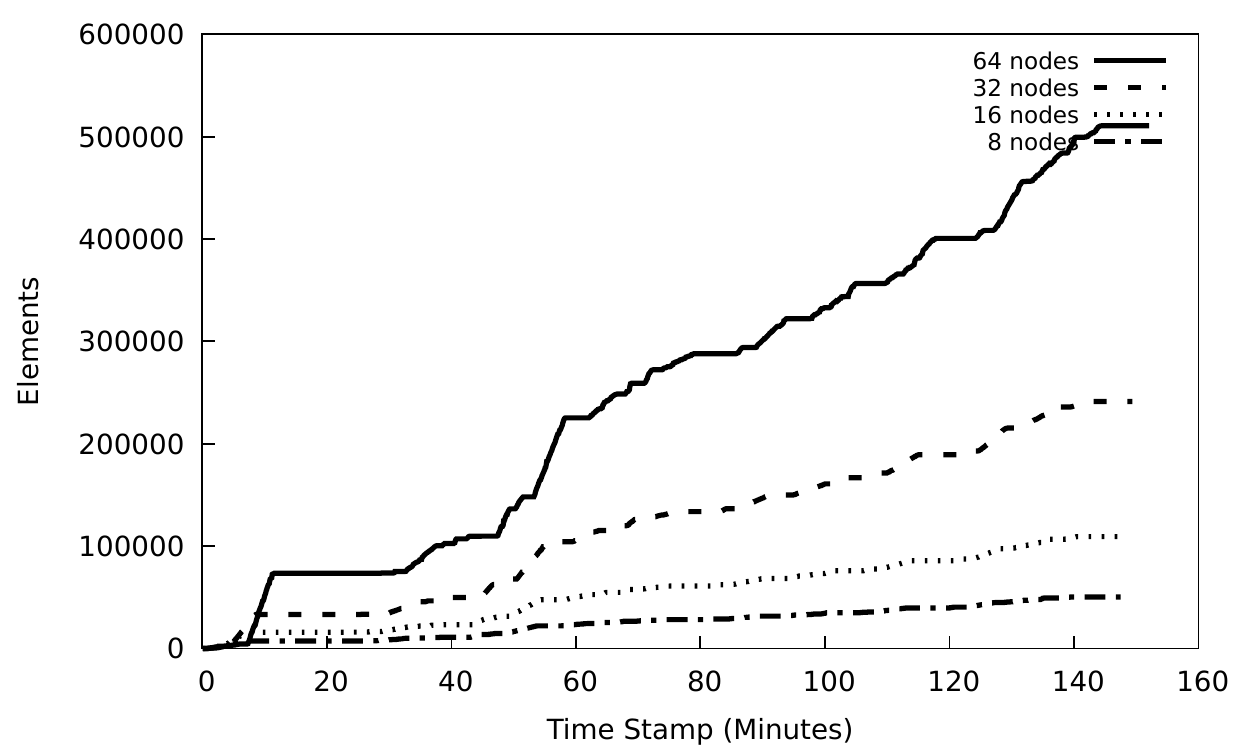}} 
 \subfloat[Max. replica count per node]{\label{fig:STORED_REPLICATIONS_MAX}\includegraphics[width = 1.7in]{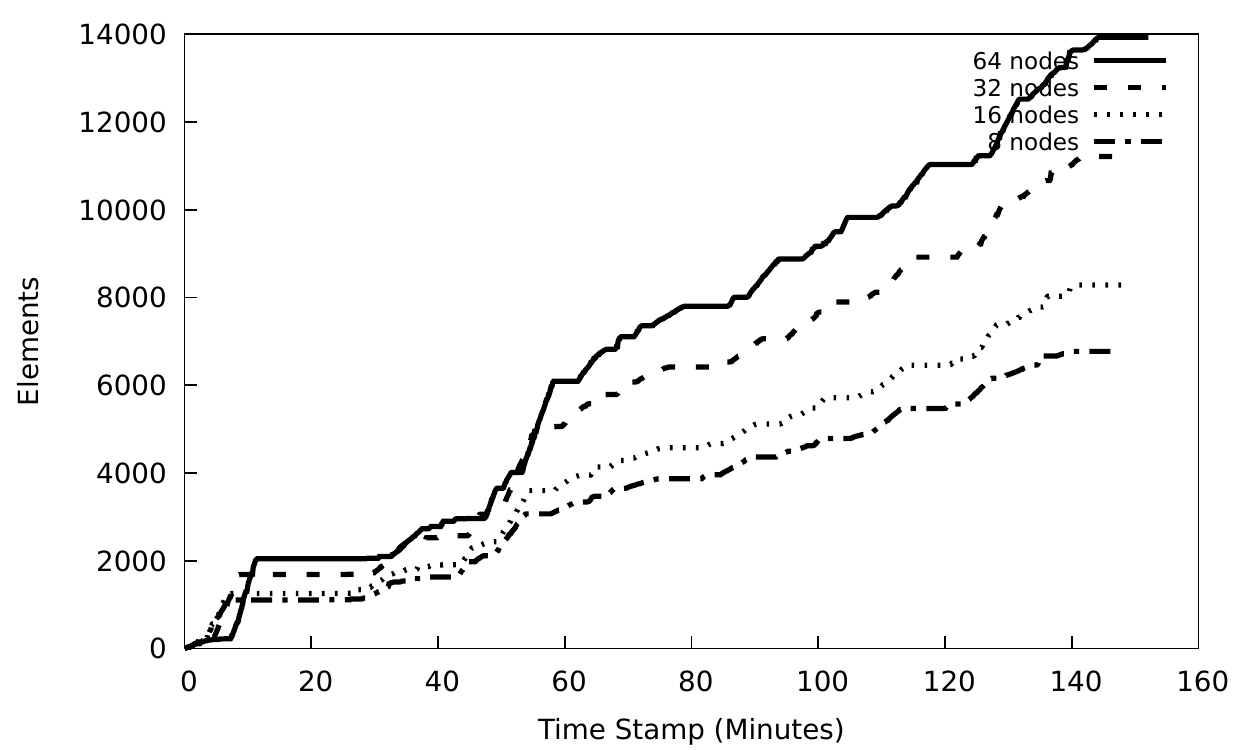}}
 \\
\vspace{-0.3cm} 
 \caption{Analysis of LibreSocial's Performance and Cost}
 \label{fig:SystemPerformance}
\end{figure*}

\section{Evaluation of performance and cost}
\label{sec:Evaluation}
In order to measure the efficacy of the system, we carry out an experiment in which all the plugins are run so as to ensure that all system functions are activated.
The quality of the system is given through the function set and provided features. 
Three aspects were considered of importance to show the general costs of the system.
These are:
\begin
 {enumerate}[label=(\alph*)]
\item
 \textbf{Network cost}: This focuses on the data rate (bytes/sec), the message rate (messages/sec) and the average hop count for lookups. 
\item
 \textbf{Storage cost}: The focus is on the retrieval time, storage time, average data stored versus actual data stored at a single node, and average number of replicas per node versus the actual number of replicas in a node, displaying the storage load. 
\item
 \textbf{Cost of security}: This looks at the impact on the data objects when either symmetric or asymmetric encryption is performed.
\end
{enumerate}

The experimental setup involves 64 LibreSocial instances interacting. 
Eight actual computers running Debian Linux run eight slave instances each with one of the computers also running an additional master instance that directed the tests and recorded the monitoring results but does not participate in the network, giving a maximum total of 64 slave nodes and a single master node.
Four sets of experiments were done, with the same workload and exponentially increasing number of 8, 16, 32 and 64 nodes.
In the first 10 minutes, the master node is initialized and then the slave nodes join in a sequential order to ensure network stability.
Thereafter, the test is conducted as shown in Table~\ref{tab:baseline_workload}, with a gap of one minute between each test. 
The specific plugin actions are reflected in the Figures~\ref{fig:PLUGIN_MESSAGING_VIEWINBOX} - \ref{fig:PLUGIN_PHOTOS_CREATEDALBUM} at the example of the count of inbox retrievals, friendship requests, group view requests and albums created.  Analogue to the workload plan, the objects are created / requested at the corresponding time of the plan. All actions were successful. 
The network and storage analysis is shown in Figures~\ref{fig:NETWORK_DATA_READ_RATE} - \ref{fig:STORED_REPLICATIONS_MAX} and for security analysis in Tables~\ref{tab:Asymmetric} and \ref{tab:Symmetric}.
A discussion of this analysis follows.

\subsection{Network} 
In the setup with 64 nodes, we observe that the total data transfer of reaches an approximate maximum of 250 MB/s in the network (see Fig. \ref{fig:NETWORK_DATA_READ_RATE}) with an maximum peak load per node at 3.9 MB/s within the first ten minutes (see Fig.~\ref{fig:PASTRY_DATA_READ_RATE}), which corresponds to the network initialization phase.
However, in general, network data rates oscillate between 0 and 150 MB/s with peaks during the network initialization phase, the file uploads and the photo uploads (see Fig. \ref{fig:PLUGIN_PHOTOS_CREATEDALBUM}). 
This corresponds to 2.4 MB/s as average peak load per node across all setups, which is acceptable. 
The number of nodes does not effect the max node load, as new nodes bring statistically more resources than they consume. 
Please note, that we present the in traffic characteristics, as the out traffic / sending transmission rates are lower than the receiving rates. 
This is due to the same homogeneous transmission load at all nodes in the work plan, i.e. homogeneous send rates, heterogeneous receive rates.  
The maximum load for sending messages is roughly half of the receiving load.  

The transmission of DDS, i.e. distributed data structures, is taking the biggest impact on the traffic, as shown in Fig.~\ref{fig:STORAGE_RETRIEVED_DDS_DATA_RATE_MAX}. 
Further traffic sources are the replication of the stored data items as well as messages. 
Messages in a network may be join messages, leave messages, maintenance messages, user messages or request and retrieval of results~\cite{KGK+08}.
The network message rate (see Fig.~\ref{fig:MESSAGES_RECEIVED_RATE} peaks at the end of the experiment with up to 13,000 messages/s, corresponding to the Wall plugin experiment. 
This is due to the retrieval of the DDS associated with the wall comments.
In general, the total message rate oscillates between 0 and 6000 messages/sec. 
Fig.~\ref{fig:MESSAGES_RECEIVED_RATE} shows the maximum per node message rate, reaching 100 message/s. Both numbers are considered low. 

%
%

We have a very low hop count of 0 to 1 in the system (see Fig.~\ref{fig:MESSAGE_HOPCOUNT_MEAN}). 
This is due to the fact that the routing table has room for $160\cdot20 = 3200$ entries, sufficient to list the maximum 64 nodes in the setup. 
Mostly, every node has information about every other node in our setup. 
As the overlay provides lookups in logarithmic time to the number of nodes in the system, only a setup with hundred thousands of nodes would increase the hop count significantly. 
Correspondingly to the low hop count, the retrieval times (see Fig. \ref{fig:RETRIEVETIME_MEAN}) and storage times (see Fig. \ref{fig:STORETIME_MEAN}) are also very low at 50 and 110 milliseconds. 
The storage needs twice the retrieval time mainly due to the creation of replicas during storage.
Both  values are considered tolerable.

\subsection{Storage} Figure~\ref{fig:NUM_STORED_LOCAL_OBJECTS} to Figure~\ref{fig:STORED_REPLICATIONS_MAX} show the storage analysis of our tests. 
The focus of the storage analysis is the storage and replication load in total and at the most loaded node. 
The overall number of unique objects is shown in Fig.~\ref{fig:NUM_STORED_LOCAL_OBJECTS}, the number of corresponding replicas is presented in Fig.~\ref{fig:NUM_STORED_REPLICATIONS}. 
As same test operations were performed at every node instance, the storage initiation load was the same for all nodes, while the actual storage node was diverse. 
As pointed out earlier, LibreSocial is designed to ensure that no single node is overwhelmed with storage requests, and that the replication requests are evenly distributed in the entire network. 
The peak load load is roughly small with 4100 unique items at the most loaded node (see Fig.~\ref{fig:NUMINSTORAGE_ELEMENTS_MAX}) and 14,000 replicated items at maximum per node (see Fig.~\ref{fig:STORED_REPLICATIONS_MAX}).
These maximum values are less than twice the average storage load of 2500 unique items in average per node and approx. 8000 replicated items in average per node. 
Thus the maximum load deviation is below 2, showing a fair load distribution. 
This shows that the additional load brought by further nodes to the system enlarges the resource pool that is used uniformly.


\subsection{Security} One of the changes between LifeSocial and LibreSocial is the public key infrastructure used.
LifeSocial implemented 1024-bit RSA algorithm~\cite{RSA78} while in LibreSocial this was changed to a 160-bit ECC algorithm~\cite{Mill86,Kob87}.
The performance of the ECC algorithm is shown in Table~\ref{tab:Asymmetric} which can be compared with the results in~\cite{GMM+09}.
In general, the overhead is much smaller than for RSA algorithm, and equally the encryption and decryption times are significantly reduced.
Also, we evaluated the AES algorithm for symmetric encryption and tabulated the results in Table~\ref{tab:Symmetric}.
The overheads are generally small as most of the objects are usually less than 1 kilobyte, with the encryption and decryption times being less than 1 millisecond.

\section{Conclusion}
\label{sec:Conclusion}
\label{Sec:Conclusion}
This paper presents LibreSocial in full, an P2P-based platform for Online Social Networks. 
The goal of the development of this OSN application is to provide a fully-distributed, secure online social network that offers with high-quality services while having practically no operational cost, despite running on unreliable, unsecure and sometimes malicious user devices.
To match the needs for such an OSN, the paper specifies technical requirements for a P2P-based OSN, and shows how LibreSocial is designed to meet these requirements.
LibreSocial is designed on a structured P2P overlay, FreePastry, with modifications for identity management and security, hence guaranteeing logarithmic routing efficiency. 
While PAST offers simple file storage, the inclusion of distributed sets, distributed linked-lists and prefix hash trees provides support for complex data such as albums, comments and inbox messages, while ensuring these have access control features, and opens up the system to the implementation of more advanced searching mechanisms such as range searches. 
The monitoring and testing plugins included in LibreSocial sets it above other systems as it allows for quality of service (QoS) monitoring using the available aggregated metrics and therefore adjustments can be made to achieve desired QoS standards.
Through the broad capabilities of the used P2P framework, LibreSocial provides simple yet powerful implementations of OSN functions, such as friends, messaging, photos, walls but also unique features such as group/forums, file hosting, voting and audio/video chat.
The modern user interface makes it compelling to use.

Selected elements of LibreSocial have been partially published before and reached very positive reaction in the community, corresponding dissertations \cite{Gra10, Amft17, Alaaridhi17, Dister18} around LibreSocial elaborate on specific elements, such as the overlay, the storage and the monitoring. 
This is the first in detail overview on the overall architecture and interdependencies of the elements. 
In general, LibreSocial offers a working solution for fully distributed, secure but also high quality social networking and is capable to further support a wide set of simple to develop applications. 

As next step, we aim to deploy LibreSocial in a beta test to gain further insights on its performance 'in the wild'. 
With this, we work on our vision to provide a feature-rich tool for secure and privacy-aware communication and interaction, that cannot be surveilled or shut down \cite{GraffiVision}, thus providing a tool for free speech in today's challenging times. 

\begin{spacing}{0.8}
 
\end{spacing}


%
%
%
%
%
%

\end{document}